\documentclass[sigconf]{acmart}
\AtBeginDocument{%
  }

\copyrightyear{2026}
\acmYear{2026}
\setcopyright{cc}
\setcctype{by}
\acmConference[CIKM '26]{Proceedings of the 35th ACM International Conference on Information and Knowledge Management}{November 07--11, 2026}{Rome, Italy}
\acmBooktitle{Proceedings of the 35th ACM International Conference on Information and Knowledge Management (CIKM '26), November 07--11, 2026, Rome, Italy}
\acmDOI{10.1145/3799682.3840999}
\acmISBN{979-8-4007-2539-5/2026/11}

\usepackage[utf8]{inputenc}
\usepackage{microtype}
\usepackage{graphicx}

\usepackage{soul}

\usepackage{hhline}
\usepackage{multirow}
\usepackage[table,xcdraw]{xcolor}
\usepackage{threeparttable}
\usepackage{algorithm}
\usepackage{algorithmic}
\usepackage{array}
\usepackage{adjustbox}
\usepackage{caption}
\usepackage{tikz}
\newcommand*\circled[1]{\tikz[baseline=(char.base)]{
            \node[shape=circle,draw,inner sep=0.5pt] (char) {#1};}}

\usepackage{colortbl}

\usepackage{hyperref}
\hypersetup{
    colorlinks=true,
    linkcolor=blue,
    filecolor=magenta,      
    urlcolor=cyan,
}
\usepackage{amsmath}
\DeclareMathOperator*{\argmax}{argmax}

\newcolumntype{L}[1]{>{\raggedright\let\newline\\\arraybackslash\hspace{0pt}}m{#1}}
\newcolumntype{C}[1]{>{\centering\let\newline\\\arraybackslash\hspace{0pt}}m{#1}}
\newcolumntype{R}[1]{>{\raggedleft\let\newline\\\arraybackslash\hspace{0pt}}m{#1}}

\usepackage{amsfonts}

\begin{document}


\title{A Dual-Expert Strategy Integrating LLMs to Mitigate Negative Transfer in Cross-Domain Sequential Recommendation}


\author{Hyeongjun Yun}
\affiliation{%
  \institution{SK Telecom, KAIST}
  \city{Seoul}
  \country{Republic of Korea}}
\email{hjyoon@sk.com}

\author{Kihyuk Song}

\affiliation{%
  \institution{SK Telecom}
  \city{Seoul}
\country{Republic of Korea}}
\email{skhskt@sk.com}

\author{Jaegul Choo}

\affiliation{%
  \institution{KAIST}
  \city{Seoul}
\country{Republic of Korea}}
\email{jchoo@kaist.ac.kr}

\author{Chung Park}
\authornote{Corresponding Author}
\affiliation{%
  \institution{SK Telecom}
  \city{Seoul}
\country{Republic of Korea}}
\email{skt.cpark@sk.com}

\renewcommand{\shortauthors}{Yun et al.}

\begin{abstract}
Cross-Domain Sequential Recommendation (CDSR) predicts the next item a user will interact with based on their historical interaction sequences across multiple domains.
Recent approaches leverage Large Language Models (LLMs) finetuned on textual representations of cross-domain user sequences to retrieve the recommended items, referred to as LLMRec.
However, LLMRec primarily models the autoregressive patterns of token-level item texts, while overlooking item-level collaborative signals. 
This semantic misalignment often leads to distorted knowledge transfer across domains—termed \textit{negative transfer}—degrading performance in the CDSR task.
To address this issue, we propose a novel LLM-based CDSR model, \textbf{DuELRec}: \underline{\textbf{D}}omain-Gated D\underline{\textbf{u}}al \underline{\textbf{E}}xperts with \underline{\textbf{L}}LMs for Cross-Domain Sequential \underline{\textbf{Rec}}ommendation.
We propose a domain-gated dual-expert framework, equipped with an item-aware attention transformation module, which aggregates textual subtokens into item-level representations and enforces block-level attention masking.
The single-domain expert restricts autoregressive attention to items within the same domain, while the cross-domain expert allows it across all domains. 
A gating mechanism adaptively fuses their outputs, using single-domain signals to reduce cross-domain noise that causes negative transfer.
Second, we introduce a dual-sampling token-to-item contrastive learning objective that allows LLMs to capture the item-level collaborative signals from both single- and cross-domains.
This is achieved by transforming token-level item texts into item-level representations and applying stochastic negative sampling from both single- and cross-domain item pools for contrastive learning.
Extensive experiments on two real-world datasets across ten domains show that our model outperforms 26 state-of-the-art methods in recommendation performance.
Its deployment in our personal assistant app led to a 47.6\% relative increase in click-through rate, demonstrating real-world impact.
Our code and appendices are available at {\color[HTML]{0037D7}\url{https://github.com/cpark88/DuELRec}}.
\end{abstract}



\maketitle

\section{Introduction}
In most real-world recommender systems, different business domains (e.g., e-commerce or short-form video platform) require a nuanced understanding of the varying interests and needs of users ~\cite{park2023cracking, park2024pacer}.
One promising approach to address these challenges is Cross-Domain Sequential Recommendation (CDSR), which predicts the next item a user will interact with by using their historical interaction sequences across multiple domains.
CDSR enhances recommendation performance by enabling a single model to effectively generalize across a user’s diverse contexts\footnote{Note that \textbf{Single-Domain Sequential Recommendation (SDSR)} focuses on predicting the next item within a specific domain by utilizing only single-domain sequences, necessitating the construction of multiple domain-specific models~\cite{kang2018self,sun2019bert4rec, zhou2020s3}.}.

There are two representative types of CDSR models: (1) ID-based Recommender (IDRec) converts items into integer IDs and creates item embedding tables for encoding ~\cite{park2023cracking, park2024pacer, cao2022contrastive}.
These IDs serve only as item indices and provide no semantic information.
(2) In contrast, Large Language Model-based Recommender (LLMRec), while promising, transforms items into textual representations and feeds them into an LLM to generate or retrieve textual descriptions (e.g., titles) of recommended items ~\cite{hou2022towards, li2023text, wang2024rethinking}.
LLMRec is able to integrate item text features and feed pre-trained knowledge from open world contexts into the recommender.
Therefore, it is scalable to other domains and demonstrates promising performance even in cold-start and cross-domain scenarios, whereas IDRec typically requires retraining from scratch to adapt to new domains or items.

However, (1) existing LLMRec models for CDSR \textbf{struggle to capture item-level collaborative signals}, namely the co-occurrence patterns between items and users within cross-domain sequences.
These models prioritize the autoregressive patterns of subtokens in item titles or descriptions, relying heavily on textual information rather than directly modeling sequential item patterns as IDRec does~\cite{kim2024large}.
(2) In particular, the transfer of misaligned signals—such as autoregressive subtoken patterns that fail to accurately reflect item-level collaborative signals—can be detrimental in CDSR tasks. 
Due to its multi-domain nature, CDSR encourages inter-domain knowledge exchange, where such signals may cause harmful interference across domains~\cite{park2023cracking, park2024pacer}.
\textbf{This leads to overall performance degradation across domains, a phenomenon known as \textit{negative transfer}}~\cite{park2024pacer}.
In other words, negative transfer is the direct consequence of such harmful interference, leading to a degradation in recommendation performance across domains.

    \begin{figure*}[tp]
    \centering
    \includegraphics[width=0.99\linewidth]{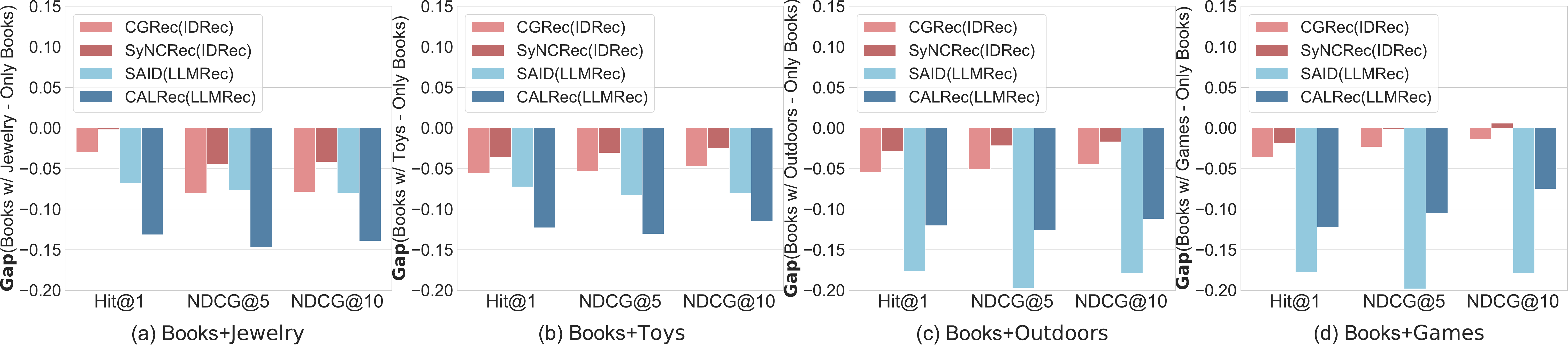}
    \caption{
The performance of the model trained on \textit{Books} alone was compared with cross-domain settings where \textit{Books} was paired with four other domains in the \textit{Amazon} dataset. 
Compared to IDRec (CGRec~\cite{park2023cracking}, SyNCRec~\cite{park2024pacer}), LLMRec (SAID~\cite{hu2024enhancing}, CALRec~\cite{li2024calrec}) experienced a more significant degradation in performance when trained jointly with other domains, relative to its performance when trained on the \textit{Books} domain alone.
    }
    \label{fig:teaser}
    \end{figure*}

As shown in Fig. \ref{fig:teaser}, to verify this experimentally, we measured the performance difference between a model trained only on the \textit{Books} domain (single-domain setting) and a model trained on the \textit{Books} domain along with four other domains (cross-domain setting) in the Amazon dataset.
Notably, in the CDSR setting, LLMRec suffers more severe performance degradation than IDRec, primarily due to its inability to capture the item-level collaborative signals accurately, which leads to the transfer of misaligned knowledge between domains.

To address these issues, we propose an LLM-based CDSR approach, called \textbf{DuELRec}: \underline{\textbf{D}}omain-gated D\underline{\textbf{u}}al \underline{\textbf{E}}xperts with \underline{\textbf{L}}LM for Cross-Domain Sequential \underline{\textbf{Rec}}ommendation, to mitigate the negative transfer problem in LLMRec.
\textbf{First}, we devise a \textit{\textbf{domain-gated dual experts}} framework to integrate item-level collaborative signals into LLMs.
At the core of this framework is an item-aware attention transformation module that explicitly restructures token-level user sequences into \textit{item-level sequences} by pooling subtokens belonging to the same item and redefining the attention mask at the block level.
The module applies distinct transformation rules to two transformer-based parallel experts: 
in the \textit{single-domain expert}, the mask is designed so that items can attend autoregressively only to previous items from the same domain, while in the \textit{cross-domain expert}, the mask allows autoregressive attention across items from all domains.
This design enforces the LLM to model dependencies at the item granularity while preserving rich subtoken semantics, thereby aligning the LLM’s attention scope with the recommendation objective.
By adaptively combining the outputs of the two experts via a gating mechanism, the model leverages single-domain signals to regulate and mitigate potential negative transfer from cross-domain interactions.
\textbf{Second}, building on the domain-gated dual sequential experts, we design a \textit{\textbf{dual-sampling token-to-item contrastive learning}} objective that enables token-level pre-trained LLMs to capture both single- and cross-domain item-level collaborative signals. 
To this end, we encode token-level item texts into item-level representations, and conduct stochastic negative sampling from both single- and cross-domain item pools to facilitate item-level contrastive learning.
This sampling strategy allows the model to effectively learn collaborative signals both within and across domains, thereby enhancing the model’s ability to differentiate between items from the same and different domains in the representation space.

To the best of our knowledge, our approach is the first ID-free LLM-based CDSR that relies solely on textual information and handles more than three domains simultaneously, whereas previous studies typically use both ID and textual features and are limited to two paired domains.
Therefore, our model eliminates the need for multiple domain-specific models with additional ID encoding look-up tables, as a single model is sufficient for cross-domain applications.
We conducted an extensive performance evaluation, comparing our model against \textbf{26 state-of-the-art IDRec and LLMRec baselines}, and demonstrated its superior performance across two real-world datasets covering ten domains.
Our model, deployed in the recommendation system of our personal assistant app, improved the click-through rate by 47.6\% (0.90\% $\rightarrow$ 1.33\%) compared to existing CDSR models, proving its effectiveness in practice.

\section{Related Work} \label{section:related_work}
\subsection{ID-based Sequential Recommendation} \label{subsection:sr}

\textbf{Single-Domain Sequential Recommendation} (SDSR) focuses on capturing temporal dynamics in user-item interactions to track evolving preferences within a domain. 
GRU-based models like GRU4Rec ~\cite{hidasi2015session} and STAMP ~\cite{liu2018stamp} predict next-item sequences, while attention-based methods such as SASRec ~\cite{kang2018self} and BERT4Rec ~\cite{sun2019bert4rec} capture short- and long-term dependencies.
Alternative architectures, including convolutional networks (NextItNet ~\cite{yuan2019simple}) and Markov Chains (TransRec ~\cite{he2017translation}), also are used in the SDSR task. 
However, SDSR requires building multiple domain-specific models, which leads to operational inefficiencies.
\noindent\textbf{Cross-Domain Recommendation} (CDR) leverages multi-domain data to improve recommendation performance, with early methods such as CMF ~\cite{singh2008relational} and CLFM ~\cite{gao2013cross} using matrix factorization to discover common patterns across multiple domains. Other methods, such as DTCDR ~\cite{zhu2019dtcdr}, DeepAPF ~\cite{yan2019deepapf}, and BiTGCF ~\cite{liu2020cross}, employed multi-task learning to capture user-item relationships in paired domains. 
\noindent\textbf{Cross-Domain Sequential Recommendation} (CDSR) extends CDR to sequential tasks spanning domains. 
Recent CDSR models such as CGRec ~\cite{park2023cracking} and SyNCRec ~\cite{park2024pacer} quantified negative transfer for each domain and penalized high-negative-transfer domains during training. 
However, their reliance on retraining from scratch for new domains or items makes them vulnerable to cold-start issues.

\subsection{LLM-based Sequential Recommendation} \label{subsection:llm_sr}

The integration of Large Language Models (LLMs) into recommendation systems has recently gained attention for addressing challenges like cross-domain transferability and cold-start problems.
UniSRec ~\cite{hou2022towards} uses LLMs and mixture-of-experts (MoE) adaptors to create isotropic and transferable item representations.
Similarly, RecFormer ~\cite{li2023text} employs a bi-directional Transformer to encode item attributes as text in an ID-free paradigm, excelling in cross-domain and zero-shot settings.
For lightweight real-world applications, E4SRec ~\cite{li2023e4srec} incorporates pluggable LLM components for scalable ID-based recommendations. 
Lite-LLMRec ~\cite{wang2024rethinking} reduces redundancy by using a hierarchical structure with separate LLMs for context encoding and inference.
Aligning item representations with LLMs has also been explored; SAID ~\cite{hu2024enhancing} maps item IDs into LLM semantic spaces via a projector module, while LLMEmb~\cite{liu2024large} incorporates collaborative signals into LLM-generated embeddings by modifying the contrastive loss.
Addressing long user behavior sequences, frameworks like ReLLa ~\cite{lin2024rella} and CALRec ~\cite{li2024calrec} improve LLM comprehension and sequential modeling through techniques like retrieval-enhanced tuning and category-specific fine-tuning strategies.
However, they are challenged by the negative transfer problem between multiple domains, where knowledge from one domain undermines performance in others.

\section{Preliminary}  \label{section:preliminary}

\noindent\textbf{Definition 1. Cross-Domain}: The domain set is represented as $\mathcal{D} = \{A, B, C, \ldots\}$, where $|\mathcal{D}| \geq 3$.
A specific domain is denoted as $d \in \mathcal{D}$. 
The set of items within a domain $d$ is expressed as $\mathcal{V}^d$. 
Consequently, the total set of items across all domains is defined as $\mathcal{V}=\cup_{d \in \mathcal{D}} \mathcal{V}^{d}$.
\ul{Since our approach retrieves from the predefined candidate set $\mathcal{V}$ based on similarity computation, it avoids the hallucination issue often seen in generative LLM-based recommender systems} (Appendix \ref{section: future_work}).

\noindent\textbf{Definition 2. Cross-Domain Sequence}: We formulate the cross-domain sequence of a user in a chronological order, $X=\{x_{1}^{C}, x_{2}^{A},...\\,x_{n}^{B} \}$, where $n$ is the length of $X$ and each interacted item $x$ is assigned a descriptive text\footnote{This sequence illustrates an example where all three domains are represented. If the arbitrary sequence consists of seven items, the illustrative and complete sequence can be represented as $X=\{x_{1}^{C}, x_{2}^{A},x_{3}^{A},x_{4}^{B},x_{5}^{C},x_{6}^{A},x_{7}^{B} \}$.}.
Each item $x$ has a corresponding attributes such as title, description and domain name.
Using this information, we set the item $x$ to be the concatenation of the description texts represented by natural languages.
The item $x$ is thus represented in a textual format (e.g., \texttt{iron man} - \texttt{action movie}).

\noindent\textbf{Problem Statement}: Our task is to predict the next item $x_{t+1}^d$ based on the historical cross-domain sequences $X_{1:t}$ up to time step $t$:

 \begin{equation}
  \begin{split}
 \label{equation:problem_statement}
  &\argmax_{x^{d}_{t+1}\in \mathcal{V}^{d}} P\Big(x^{d}_{t+1} \Big\vert X_{1:t} \Big), 
 \end{split}
 \end{equation} 
where $P\Big(x^{d}_{t+1} \Big\vert X_{1:t} \Big)$ represents the likelihood that the target item $x^{d}_{t+1}$ in domain $d$ will be interacted with, given item sequences $X_{1:t}$.

\section{Model}  \label{section:model}

\begin{figure*}[tp]
\centering
\includegraphics[width=1.0\linewidth]{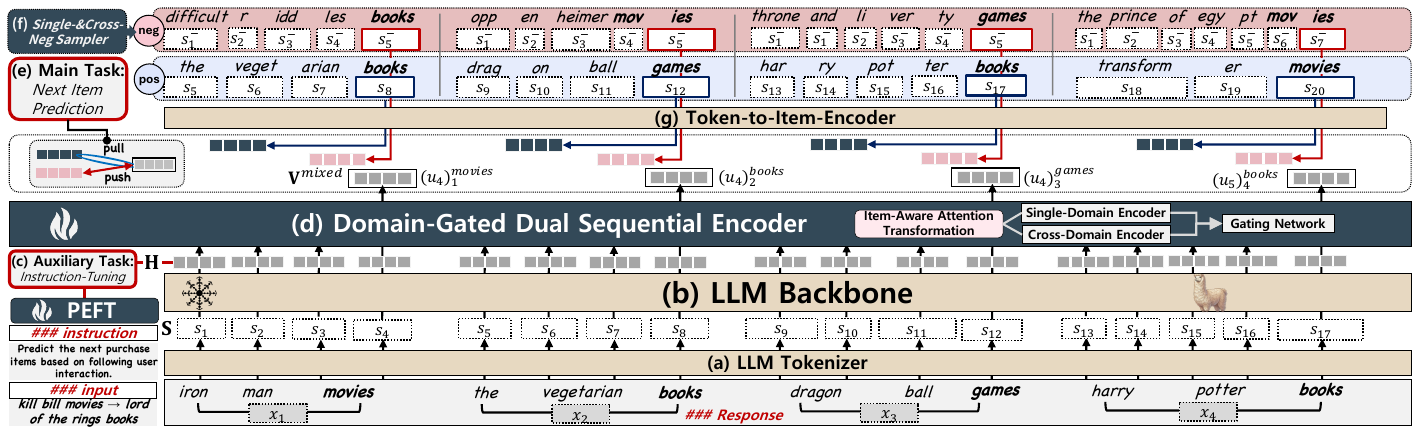}
\caption{
We illustrate our model with three domains \textit{Movies}, \textit{Books}, and \textit{Games}.
The item-level sequence has a total length of $n=4$, while the subtoken-level sequence has a length of $L=17$.
}
\label{fig:overall_model}
\end{figure*}

\subsection{LLM Tokenizer (Fig. \ref{fig:overall_model}a)} \label{subsection:tokenizer} 
The model input is the textual representation of the cross-domain user sequence $X$, which is fed into the tokenizer $\mathcal{I}$ and subsequently transformed into a subtoken-level sequence:

 \begin{equation}
 \begin{split}
 \label{equation:input_sequence_reformulation2}
  &S=\mathcal{I}(X) \\
  & \;\;= \{(s_{1},...,s_{l_{x_1}})_{1}^{C}, (s_{1},...,s_{l_{x_2}})_{2}^{A},...,(s_{1}, ... ,s_{l_{x_n}})_{n}^{B}\},
 \end{split}
 \end{equation}

where $l_x$ is the number of subtokens in the item $x$ from the tokenizer $\mathcal{I}$ and the total length of the subtoken level sequence $S$ is $L=\sum_{i=1}^{n} l_{x_{i}}$.
Using the subtoken level sequence $S$ (Eq.\ref{equation:input_sequence_reformulation2}), which is output by the tokenizer $\mathcal{I}$ with $X$, we can derive the sequence of subtoken embeddings $\textbf{E}\in \mathbb{R}^{L \times m}$ as:

 \begin{equation}
 \begin{split}
 \label{equation:embedding_sequence_reformulation}
  &\textbf{E}= \texttt{emb}(S) \\
  &\;\;= \{(e_1,...,e_{l_{x_{1}}})_{1}^{C}, (e_1,...,e_{l_{x_{2}}})_{2}^{A},..., (e_1,...,e_{l_{x_{n}}})_{n}^{B}\},
 \end{split}
 \end{equation}
where \texttt{emb} is the embedding layer of the LLM, $e$ is the embedding  of the subtoken $s$, and $m$ is the dimension size of \texttt{emb}.

\subsection{LLM backbone (Fig. \ref{fig:overall_model}b)} \label{subsection:llm_backbone}
The LLM backbone takes the sequence of subtoken embeddings \textbf{E} as input and produces the hidden representation $\textbf{H}\in \mathbb{R}^{L \times m}$ as follows:

 \begin{equation}
 \begin{split}
 \label{equation:hidden_sequence_reformulation}
  &\textbf{H}=\texttt{LLM}(\textbf{E})\\
  &\;\;\;=\{(h_1,...,h_{l_{x_{1}}})_{1}^{C}, (h_1,...,h_{l_{x_{2}}})_{2}^{A},..., (h_1,...,h_{l_{x_{n}}})_{n}^{B}\},
 \end{split}
 \end{equation}
where \texttt{LLM} is the LLM backbone before the output projection layer, and $h$ is the representation vector of the subtoken $s$ from the LLM backbone.

\subsection{Instruction-Tuning (Auxiliary Task; Fig. \ref{fig:overall_model}c)} \label{subsection:sft}
In this \textbf{auxiliary task}, our goal is to adapt the LLM backbone via instruction tuning tailored to recommendation tasks, enabling the model to effectively encode user behavior sequences. 
Importantly, this tuning is not intended for generating recommended items, but for structuring the LLM’s representation space around user-item interaction patterns.
The instruction tuning is performed using the \textit{Alpaca} template and a Parameter-Efficient Fine-Tuning (PEFT) method, as shown in Fig. \ref{fig:overall_model}c.

To simplify the explanation, we omit the subscripts for domain and time step for all subtokens in this section (i.e., we flatten the sequences). 
Accordingly, we represent the sequences \textit{S} (Eq. \ref{equation:input_sequence_reformulation2}) and \textbf{H} (Eq. \ref{equation:hidden_sequence_reformulation}) as $\textit{S} = \{s_1, s_2, \ldots, s_L\}$ and $\textbf{H} = \{h_1, h_2, \ldots, h_L\}$, respectively.
Based on these notations, we describe the task as follows:

 \begin{equation}
 \begin{split}
 \label{equation:clm}
  &\mathcal{L}_{inst}(\theta, \Phi) = - \frac{1}{L} \sum_{i=1}^{L} \log P_{\theta,\Phi}(s_{i+1} \mid s_{<i+1}), \\
  & P_{\theta,\Phi}(s_{i+1}\mid s_{<i+1})=\texttt{softmax}\big(\texttt{LM head}(h_{i+1})\big),
 \end{split}
 \end{equation}
where \texttt{LM head} denotes the output projection layer of the LLM backbone; $\theta$ represents its frozen parameters, and $\Phi$ denotes the parameters introduced by the PEFT method, such as low-rank matrices in LoRA~\cite{hu2021lora}.
The instruction-tuned LLM serves not merely as a text encoder, but as a task-adapted component that contributes meaningfully to the recommendation performance (Section \ref{subsection: discussion_model_variants}).

    \begin{figure}[htbp]
    \centering
    \includegraphics[width=0.999\linewidth]{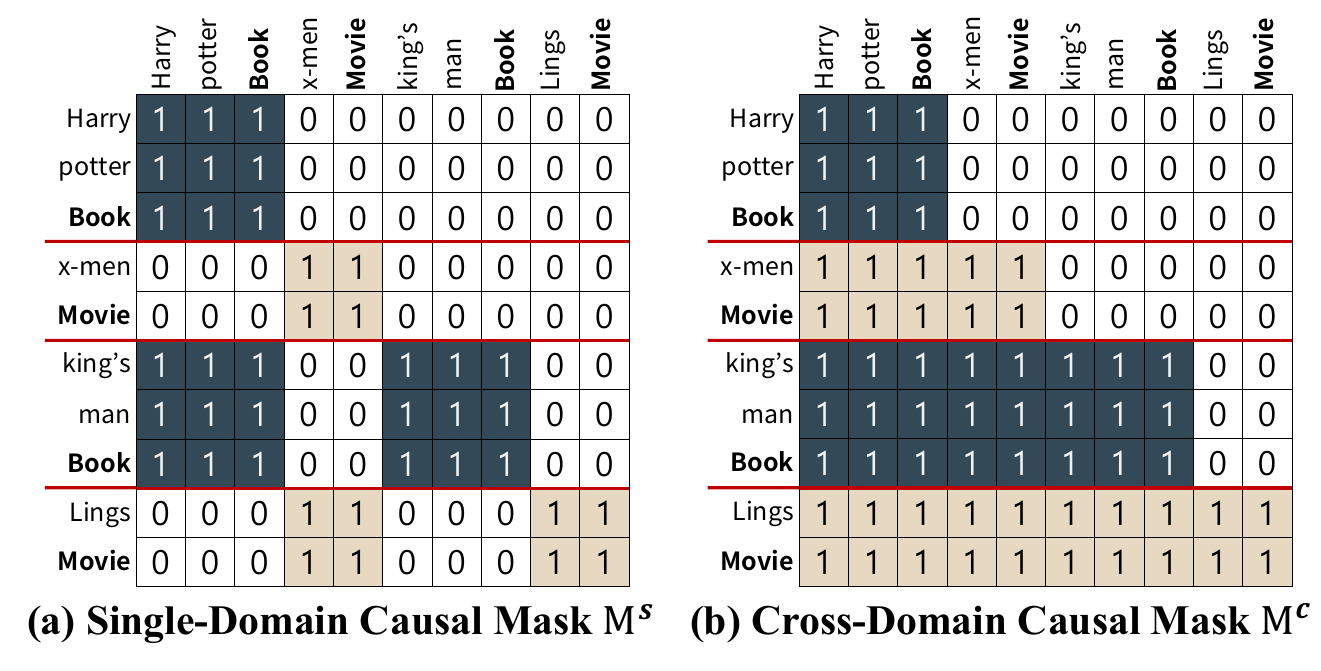}
    \caption{
    Illustration of the (a) single- and (b) cross-domain causal masks in the item-aware attention transformation.
    }
    \label{fig:llm_mask}
    \end{figure}

\subsection{Domain-Gated Dual Experts 
(Fig. \ref{fig:overall_model}d)} \label{subsection:dgde}
Learning the sequential pattern of a user's subtoken-level sequence by an LLM alone, as in previous LLMRec studies~\cite{li2023text, kim2024large, li2024calrec}, is insufficient to capture the item-level co-occurrence, as the subtoken-level modeling struggles to detect collaborative signals from user-item interactions.
To address this limitation, we introduced a novel transformer-based expert module on top of the LLM backbone to effectively learn these item-level collaborative signals in single- and cross-domain perspectives.

\subsubsection{Item-Aware Attention Transformation}
A single-domain sequence refers to user sequences composed exclusively of items from a specific domain. 
The collaborative signal within such a sequence reflects the pure characteristics of that domain, free from interference from other domains, and thus does not experience negative transfer.
However, it also forgoes the potential benefits of positive transfer across domains.
In contrast, the collaborative signal in cross-domain sequences is influenced by positive or negative transfer effects between domains.
Therefore, optimizing the use of both collaborative signals is crucial to mitigate negative transfer between domains~\cite{liu2024beyond}.

For this, we build two transformer-based experts~\cite{vaswani2017attention} to capture the item-level collaborative signals of single- and cross-domain sequences, respectively.
For these two experts, we apply the \textbf{item-aware attention transformation} for single- and cross-domain sequences.
As shown in Fig. \ref{fig:llm_mask}a, for single-domain sequences, a causal mask $\texttt{M}^{s}$ is applied in the transformer only between item-level interactions within the same domain, ensuring that attention is restricted to items from that specific domain. 
Attention within an item is applied to all of its subtokens, allowing each subtoken to attend to any other subtoken within the same item.
In contrast, for cross-domain sequences, attention is applied to all items, but with an item-level causal mask $\texttt{M}^{c}$, ensuring that sequential dependencies are preserved across domains (Fig. \ref{fig:llm_mask}b).
Specifically, the single- ($\texttt{EXPERT}^{s}$) and cross-domain experts ($\texttt{EXPERT}^{c}$) take the output of LLM backbone $\textbf{H}$ (Eq. \ref{equation:hidden_sequence_reformulation}) and causal masks $\texttt{M}^{s}$, $\texttt{M}^{c}$ as inputs, and derive the item-level representation as follows:

 \begin{equation}
 \begin{split}
 \label{equation:encoder_reformulation}
  &\textbf{U}^{s}=\texttt{EXPERT}^{s}(\textbf{H},\texttt{M}^{s}) \\
  & \;\;\;\;=\{(u_1,...,u_{l_{x_{1}}})_{1}^{C}, (u_1,...,u_{l_{x_{2}}})_{2}^{A},..., 
   (u_1,...,u_{l_{x_{n}}})_{n}^{B}\}^{s},  \\
  &\textbf{U}^{c}=\texttt{EXPERT}^{c}(\textbf{H},\texttt{M}^{c})\\
  &\;\;\;\;=\{(u_1,...,u_{l_{x_{1}}})_{1}^{C}, (u_1,...,u_{l_{x_{2}}})_{2}^{A},..., (u_1,...,u_{l_{x_{n}}})_{n}^{B}\}^{c}, 
 \end{split}
 \end{equation}
where $\textbf{U}^{s}, \textbf{U}^{c} \in \mathbb{R}^{L \times m} $ are the output representations of single- and cross-domain sequences of two transformer-based experts, and $L=\sum_{i=1}^{n} l_{x_{i}}$ is the total length of the $\textbf{U}^{s}$ and $\textbf{U}^{c}$.

We use only the last element of each item as the item-level representation, since it contains the semantic information and the collaborative signal of the given item.
Therefore, we can describe it from $\textbf{U}^{s}, \textbf{U}^{c} \in \mathbb{R}^{L \times d}$ in Eq. \ref{equation:encoder_reformulation}, as $\textbf{V}^s, \textbf{V}^c \in \mathbb{R}^{n \times m}$ as follows:

 \begin{equation}
 \begin{split}
 \label{equation:encoder_reformulation_last}
  &\textbf{V}^s = \{(u_{l_{x_{1}}})_{1}^{C},(u_{l_{x_{2}}})_{2}^{A},...,(u_{l_{x_{n}}})_{n}^{B}\}^{s} \\
  &\textbf{V}^c = \{(u_{l_{x_{1}}})_{1}^{C},(u_{l_{x_{2}}})_{2}^{A},...,(u_{l_{x_{n}}})_{n}^{B}\}^{c}.
 \end{split}
 \end{equation}

\subsubsection{Item-Wise Gating Network}
We then developed a gating network to compute the optimal weighting between the two representations at the \textbf{item-level} (i.e., $\textbf{V}^s$ and $\textbf{V}^c$), allowing the model to leverage single-domain signals to regulate and mitigate potential negative transfer from cross-domain interactions.
The gating network~\cite{shazeer2017outrageously}, $g$, which computes the final representation $\textbf{V}^{mixed} \in \mathbb{R}^{n \times m}$ as follows:

 \begin{equation}
 \begin{split}
 \label{equation:gating_module}
  & \textbf{V}^{mixed} = g( [\textbf{V}^{s}+\textbf{V}^{c}] )[: \;,0] \; \textbf{V}^{s} + g( [\textbf{V}^{s}+\textbf{V}^{c}] )[: \;,1] \; \textbf{V}^{c} \\
  &\;\;\;\;\;\;\;\;\;\;\;=\{(u_{l_{x_{1}}})_{1}^{C}, (u_{l_{x_{2}}})_{2}^{A},..., 
   (u_{l_{x_{n}}})_{n}^{B}\}^{mixed}, \\
  & g(\textbf{Y}) = \texttt{softmax}\big(W^{1}\textbf{Y} + (\texttt{StdNorm}*\texttt{SoftPlus}(W^{2}\textbf{Y}))\big), \\
 \end{split}
 \end{equation}
where $\texttt{softmax}$, $\texttt{StdNorm}$, and $\texttt{SoftPlus}$ 
denote the softmax function, the noise from the standard normal distribution, and the softplus activation function, respectively. 
In addition, $W^1, W^2 \in \mathbb{R}^{m \times 2}$ are the trainable fully-connected layers.
Therefore, the gating network $g$ produces the adaptive weighting tensor shaped as $\mathbb{R}^{n \times 2}$, and balances the $\textbf{V}^s$ and $\textbf{V}^c$ on a per-item basis to reduce the negative transfer between domains.

We theoretically show that the gating mechanism mitigates inter-domain negative transfer via a bias-variance decomposition.
Let $\mathbf{s}_i$ be the true item representation.
The single-domain expert provides an unbiased estimate $\mathbf{v}^{s}_i = \mathbf{s}_i + \boldsymbol{\varepsilon}^{s}_i$ with variance $\sigma_s^2$, while the cross-domain expert yields an output affected by negative transfer, $\mathbf{v}^{c}_i = \mathbf{s}_i + \mathbf{b}_i + \boldsymbol{\varepsilon}^{c}_i$, where $\mathbf{b}_i$ denotes the cross-domain interference term and $\sigma_c^2$ the noise variance.
The mixed representation is computed as $\mathbf{v}^{\textit{mixed}}_i = g_i \mathbf{v}^{s}_i + (1 - g_i)\mathbf{v}^{c}_i$, with expected squared error:
\[
\mathbb{E}\left[\|\mathbf{v}^{\textit{mixed}}_i - \mathbf{s}_i\|^2\right] = g_i^2 \sigma_s^2 + (1 - g_i)^2 (\|\mathbf{b}_i\|^2 + \sigma_c^2).
\]
Let $A_i:=\|\mathbf{b}_i\|^2+\sigma_c^2$ and $f(g_i):=g_i^2\sigma_s^2+(1-g_i)^2A_i$.
Since $\frac{d^2 f}{d g_i^2}=2(\sigma_s^2+A_i)>0$, $f$ is strictly convex, and the unique minimizer is:
\[
g_i^\star = \frac{\|\mathbf{b}_i\|^2+\sigma_c^2}{\sigma_s^2+\|\mathbf{b}_i\|^2+\sigma_c^2}.
\]
\[
g_i^\star =
\begin{cases}
< 0.5, & \text{if } \|\mathbf{b}_i\|^2 + \sigma_c^2 < \sigma_s^2 \quad \text{(cross-domain expert preferred)} \\
> 0.5, & \text{if } \|\mathbf{b}_i\|^2 + \sigma_c^2 > \sigma_s^2 \quad \text{(single-domain expert preferred)}
\end{cases}
\]
The gating mechanism thus adaptively emphasizes the more reliable expert based on the relative bias and variance, providing a principled way to suppress unreliable signals and reduce negative transfer in a domain-aware manner.
When the cross-domain expert suffers from a large combined bias–variance, the optimal weight shifts toward the single-domain expert (Fig.~\ref{fig:gating_weight}).

\subsection{Dual-Sampling Token-to-Item Contrastive Learning (Main Task; Fig. \ref{fig:overall_model}e)} \label{subsection:nip}
In addition to identifying correlations among similar items, we also investigate contrasts between dissimilar items.
The pairwise ranking loss is adopted by conventional recommenders for personalized ranking, i.e., learning users' preferences for some items over the others.
This training task allows LLMs to understand the collaborative signal on an item-by-item basis.
We optimize our model by leveraging pairwise ranking loss on a cross-domain sequence $X_{1:n}$ and its expected next item $x^{d}_{n+1}$ as follows:

 \begin{equation}
 \begin{split}
 \label{equation:cross_objective}
  &q_{t} = \text{log}\sigma \Big(P(x_{t+1}^{d}=x^{+} | X_{1:t} ) - P(x_{t+1}^{d}=x^{-} | X_{1:t})\Big), \\ &\mathcal{L}_{nip} = \sum_{t=1}^{n} q_{t},
 \end{split}
 \end{equation}
where $\mathcal{L}_{nip}$ is the next item prediction loss, $\sigma$ represents the sigmoid function, and $x^{+}$ denotes the ground-truth item, which is paired with a negative sample $x^{-}$ from the item vocabulary.

\subsubsection{Token-to-Item Embedding Encoder}
$P(x_{t+1}^{d}=x | X_{1:t}) $ in Eq. \ref{equation:cross_objective} is calculated as follows: 

 \begin{equation}
 \begin{split}
 \label{equation:prob_enc}
    &P(x_{t+1}^{d}=x | X_{1:t}) = \sigma \Big( u_{l_{x_t}}^{\texttt{T}} \cdot \texttt{ItemEmb}(x) \Big),
 \end{split}
 \end{equation}
where $\cdot$ is the dot-product operation, and $u_{l_{x_t}}$ is the sequence representation until $t$ time-step (Eq. \ref{equation:gating_module}) and $\texttt{ItemEmb}$ is the item-level embedding of item $x$ (Fig. \ref{fig:overall_model}g).
\texttt{ItemEmb}($x$) is the last representation of LLM-backbone for a specific item $x$.
It can be shared with the LLM of Section \ref{subsection:llm_backbone}.
If the subtoken-level sequence of item $x$ is $[s_{1},...,s_{l_{x}}]$, then $\texttt{ItemEmb}(x)$ can be described as:

 \begin{equation}
 \begin{split}
 \label{equation:item_encoder}
    &\texttt{ItemEmb}(x) = \texttt{LLM}([s_{1},...,s_{l_{x}}])_{l_x}.
 \end{split}
 \end{equation}
Alternatively, \texttt{ItemEmb} can be obtained from a separate trainable layer (e.g., Fully-Connected Layer) instead of the LLM backbone.

\subsubsection{Stochastic Single- and Cross-Domain Negative Sampler} 
Prior CDSR studies~\cite{cao2022contrastive, sun2019bert4rec} typically sample negative samples randomly from the same domain as the ground-truth item.
We introduce a contrastive learning-based negative sampling method that extends sampling from a single- to cross-domain scenarios (Fig. \ref{fig:overall_model}f). 
By randomly selecting negative samples both within and across domains at the item level, our method enables the model to learn collaborative signals not only within a domain but also across domains.
The model is thereby enhanced in its ability to distinguish domain-specific from cross-domain representations.

Suppose that $\mathcal{U}(S)$ denotes a uniform distribution within a set $S$.
For single-domain negative sampling, the $x^{-}$ is obtained by $x^{-} \sim \mathcal{U}(\mathcal{V}^{d})$, where $\mathcal{V}^{d}$ is the set of items specific to the domain $d$.
In addition, for cross-domain sampling, the $x^{-}$ is extracted by $x^{-} \sim \mathcal{U}(\mathcal{V} \texttt{\textbackslash} \mathcal{V}^{d})$, where $\mathcal{V}=\cup_{d \in \mathcal{D}} \mathcal{V}^{d}$.
Since the negative sample is drawn from either the single- or cross-domain pool with a uniform probability $\alpha$, the negative sample $x^{-}$ is described as:

 \begin{equation} \label{equation:negative_sampling}
  x^{-} \sim 
  \begin{cases}
  \;\;\mathcal{U}(\mathcal{V}^{d}), \;\;\;\;\;\;\;\;\;\;\;\;\; \text{if}\;\; \alpha > p, \\
  \;\;\mathcal{U}(\mathcal{V}\texttt{\textbackslash} \mathcal{V}^{d})\;, \;\;\;\;\;\;\; \text{if}\;\; \alpha \leq p, 

  \end{cases}
  \end{equation}
where $p$ is the hyperparameter that controls the degree of cross-domain sampling.
Since larger $p$ values increase cross-domain negative sampling, we varied $p$ from $0.0$ to $0.5$ and observed a robust increasing trend in performance. 
This indicates that including a moderate proportion of cross-domain negatives yields robust and superior overall performance, with detailed results provided in Section~\ref{subsection:hyper_ablation}.
Note that prior studies~\cite{cao2022contrastive, sun2019bert4rec} correspond to the case of $p=0$, where negative samples are drawn exclusively from the same single domain as the ground-truth item, and we empirically observed that this setting yielded the lowest performance.

\subsection{Training and Evaluation} \label{subsection:train_eval}
The total training loss function for our model is defined as follows:

 \begin{equation}
 \label{equation:total_training}
 \begin{split}
  &\mathcal{L} = \eta  \mathcal{L}_{inst}  + (1-\eta) \mathcal{L}_{nip},
 \end{split}
 \end{equation}
where $\eta$ is the harmonic factor between $\mathcal{L}_{inst}$ (Eq. \ref{equation:clm}) and $\mathcal{L}_{nip}$ (Eq. \ref{equation:cross_objective}).
In the evaluation stage, we only used the last representations of the sequence to make predictions.
For example, using the latest representations $u_{l_{x_n}}^{mixed}$ (Eq. \ref{equation:gating_module}), the next recommended item during evaluation is determined by selecting the one with the highest prediction score in domain $d$:

 \begin{equation}
  \begin{split}
 \label{equation:evaluation}
  &\argmax_{x^d\in \mathcal{V}^{d}} \big( (u_{l_{x_n}}^{mixed})^{\texttt{T}} \cdot \texttt{ItemEmb}(x^d) \big),
 \end{split}
 \end{equation} 
where $\mathcal{V}^d$ is the item set of domain $d$ and $\texttt{ItemEmb}$ is the item-level embedding encoder (Eq.~\ref{equation:item_encoder}).
The inference and deployment process in real-world application are described in Appendix \ref{section: inf_and_dep}.

\section{Experiments} \label{section:experiments}
The aim of the experiments is to answer the following questions:

\noindent\setlength{\fboxsep}{0pt}\colorbox[HTML]{DAE8FC}{\textbf{(RQ1)}}: Can our model perform better than state-of-the-art baselines in real-world scenarios involving multiple domains simultaneously, especially \textbf{more than three}?

\noindent\setlength{\fboxsep}{0pt}\colorbox[HTML]{DAE8FC}{\textbf{(RQ2)}}: Is our model able to effectively mitigate \textbf{negative transfer} in all domains of the CDSR task?

\noindent\setlength{\fboxsep}{0pt}\colorbox[HTML]{DAE8FC}{\textbf{(RQ3)}}: What is the performance of the model under \textbf{cold and warm item/user scenarios in CDSR task}?

\noindent\setlength{\fboxsep}{0pt}\colorbox[HTML]{DAE8FC}{\textbf{(RQ4)}}: How do \textbf{different components} of our model affect its overall performance?

\noindent\setlength{\fboxsep}{0pt}\colorbox[HTML]{DAE8FC}{\textbf{(RQ5)}}: How sensitive is the model performance to the hyperparameter $p$ in the \textit{Dual-Sampling Token-to-Item Contrastive Learning}?

\noindent\setlength{\fboxsep}{0pt}\colorbox[HTML]{DAE8FC}{\textbf{(RQ6)}}: Does our proposed method achieve competitive \textbf{time complexity} in both training and inference when compared to baselines?

\noindent\setlength{\fboxsep}{0pt}\colorbox[HTML]{DAE8FC}{\textbf{(RQ7)}}: How effectively does the model perform when deployed in an \textbf{online environment}?

\begin{table}[htbp]
\caption{Statistics of datasets.
}
\label{tab:data_summary}
    \renewcommand{\arraystretch}{0.85}
\begin{adjustbox}{width=\columnwidth,center}
\begin{tabular}{c|ccccc}
\toprule
\rowcolor[HTML]{EFEFEF} 
\textbf{Dataset}                  & \textbf{\#Users}          & \textbf{Domain} & \textbf{\#Items} & \textbf{\#Interactions} & \textbf{Sparsity} \\ \hline
                                  &                           & Books           & 425,985          & 1,422,676               & 99.98\%           \\
                                  &                           & Games           & 29,013           & 292,891                 & 99.97\%           \\
                                  &                           & Jewelry        & 290,804          & 947,417                 & 99.98\%           \\
                                  &                           & Outdoors          & 133,066          & 541,717                 & 99.99\%           \\
\multirow{-5}{*}{\textbf{Amazon}} & \multirow{-5}{*}{105,364} & Toys            & 121,559          & 575,449                 & 99.98\%           \\ \hline
                                  &                           & Web-View         & 9,192            & 36,716,069              & 96.01\%           \\
                                  &                           & Call-Log       & 3,301            & 3,038,385               & 99.08\%           \\
                                  &                           & PoI             & 549              & 1,892,868               & 96.44\%           \\
                                  &                           & Benefits          & 72               & 401,065                 & 92.08\%           \\
\multirow{-5}{*}{\textbf{Telco}}  & \multirow{-5}{*}{99,936}  & Shopping        & 714              & 230,233                 & 99.60\%           \\ 
\bottomrule
\end{tabular}
\end{adjustbox}
\end{table}

\subsection{Datasets} \label{subsection:datasets}
\textbf{(1) Amazon Review Dataset}~\cite{mcauley2015image} was utilized in our experiments, encompassing five domains: \textit{Books}, \textit{Video and Games}, \textit{Clothing Shoes and Jewelry}, \textit{Sports and Outdoors}, and \textit{Toys and Games} (Table \ref{tab:data_summary}).  
These domains are referred to as \textbf{\textit{Books}}, \textbf{\textit{Games}}, \textbf{\textit{Jewelry}}, \textbf{\textit{Outdoors}}, and \textbf{\textit{Toys}} for brevity.

\noindent \textbf{(2) Telco Behavior Dataset} was compiled from user logs across various real-world applications managed by a leading global Telcomunications company. The dataset includes data from customers who provided consent for its collection and analysis. It spans five distinct domains: \textit{Web View history} (\textbf{\textit{Web-View}}), \textit{Call Record} (\textbf{\textit{Call-Log}}), \textit{Navigation PoI Search history} (\textbf{\textit{PoI}}), \textit{Subscriber Benefits Program} (\textbf{\textit{Benefits}}), and \textit{e-commerce Purchase history} (\textbf{\textit{Shopping}}).

\begin{table*}[t] \footnotesize
\centering
\caption{Model performance comparison on \textit{\textbf{Amazon}}, with the top two methods highlighted in \setlength{\fboxsep}{0pt}\colorbox[HTML]{DAE8FC}{blue} and \setlength{\fboxsep}{0pt}\colorbox[HTML]{FFCCC9}{red}, respectively.
}
    \renewcommand{\arraystretch}{1.05}
\label{tab:result_1}
\setlength{\tabcolsep}{0.79pt}
\begin{tabular}{cc|ccccc|ccccc|ccccc|ccccc|ccccc}
\toprule
\multicolumn{2}{c|}{\cellcolor[HTML]{EFEFEF}\textbf{Domain}}                                                                                & \multicolumn{5}{c|}{\cellcolor[HTML]{EFEFEF}\textbf{Books}}                                                       & \multicolumn{5}{c|}{\cellcolor[HTML]{EFEFEF}\textbf{Games}}                                                       & \multicolumn{5}{c|}{\cellcolor[HTML]{EFEFEF}\textbf{Jewelry}}                                                    & \multicolumn{5}{c|}{\cellcolor[HTML]{EFEFEF}\textbf{Outdoors}}                                                      & \multicolumn{5}{c}{\cellcolor[HTML]{EFEFEF}\textbf{Toys}}                                                        \\ \hline\hline
\multicolumn{1}{c|}{}                                                                          &                                   & \multicolumn{2}{c}{HR}                      & \multicolumn{2}{c}{NDCG}                    & MRR                   & \multicolumn{2}{c}{HR}                      & \multicolumn{2}{c}{NDCG}                    & MRR                   & \multicolumn{2}{c}{HR}                      & \multicolumn{2}{c}{NDCG}                    & MRR                   & \multicolumn{2}{c}{HR}                      & \multicolumn{2}{c}{NDCG}                    & MRR                   & \multicolumn{2}{c}{HR}                      & \multicolumn{2}{c}{NDCG}                    & MRR                  \\ \cline{3-27} 
\multicolumn{1}{c|}{\multirow{-2}{*}{\textbf{Types}}}                                                   & \multirow{-2}{*}{\textbf{Models}} & @5                   & @10                  & @5                   & @10                  & @10                   & @5                   & @10                  & @5                   & @10                  & @10                   & @5                   & @10                  & @5                   & @10                  & @10                   & @5                   & @10                  & @5                   & @10                  & @10                   & @5                   & @10                  & @5                   & @10                  & @10                  \\ \hline
\multicolumn{1}{c|}{}                                                                          & GRU4Rec                           & 0.257                & 0.345                & 0.186                & 0.215                & 0.175                 & 0.558                & 0.662                & 0.435                & 0.469                & 0.408                 & 0.288                & 0.391                & 0.203                & 0.236                & 0.188                 & 0.320                & 0.432                & 0.226                & 0.262                & 0.210                 & 0.310                & 0.414                & 0.217                & 0.251                & 0.200                \\
\multicolumn{1}{c|}{}                                                                          & SASRec                            & 0.268                & 0.356                & 0.190                & 0.219                & 0.177                 & 0.560                & 0.659                & 0.443                & 0.473                & 0.417                 & 0.262                & 0.364                & 0.189                & 0.222                & 0.178                 & 0.322                & 0.425                & 0.232                & 0.265                & 0.216                & 0.328               & 0.439                & 0.235                & 0.271                & 0.219                \\

\multicolumn{1}{c|}{}                                                                          & BERT4Rec                          & 0.210                     &0.299                      &0.148                      &0.177                      &0.140                       &0.563                      &0.673                      &0.434                      &0.473                      &0.411                       &0.262                      &0.366                      &0.184                      &0.217                      &0.172                       &0.306                      &0.422                      &0.215                      &0.252                      &0.200                       &0.295                      &0.415                      &0.206                      &0.245                      &0.193                      \\
\multicolumn{1}{c|}{}                                                                          & S3Rec                             &0.210                      &  0.300                    &0.146                      &0.175                      &0.137                       &0.324                      &0.387                      &0.253                      &0.274                      &0.238                       &0.208                      &0.286                      &0.147                      &0.172                      &0.137                       &0.264                      &0.339                      &0.196                      &0.220                      &0.183                       &0.247                      &0.331                      &0.177                      &0.204                      &0.164                      \\
\multicolumn{1}{c|}{}                                                                          & NextItNet                         &0.237                      &0.327                      &0.166                      &0.195                      &0.155                       &0.533                      &0.671                      &0.388                      &0.433                      &0.359                       &0.267                      &0.379                      &0.183                      &0.219                      &0.170                       &0.305                      &0.426                      &0.210                      &0.249                      &0.195                       &0.304                      &0.425                      &0.210                      &0.249                      &0.196                      \\
\multicolumn{1}{c|}{}                                                                          & SINE                              &0.265                      &0.365                      &0.192                      &0.224                      &0.181                       &0.661                      &0.760                      &0.527                      &0.559                      &0.496                       &0.277                      &0.396                      &0.191                      &0.230                      &0.179                       &0.352                      &0.471                      &0.248                      &0.286                      &0.229                       &\cellcolor[HTML]{FFCCC9}0.361                      &0.484                      &\cellcolor[HTML]{FFCCC9}0.255                      &\cellcolor[HTML]{FFCCC9}0.294                      &0.236                      \\
\multicolumn{1}{c|}{}                                                                          & STAMP                             &0.251                     &0.351                      &0.179                      &0.212                      &0.169                       &0.539                      &0.623                      &0.425                      &0.453                      &0.399                       &0.278                      &0.381                      &0.196                      &0.229                      &0.183                       &0.316                      &0.425                      &0.222                      &0.258                      &0.207                       &0.317                      &0.430                      &0.223                      &0.260                      &0.207                      \\
\multicolumn{1}{c|}{}                                                                          & TransRec                          &0.262                      &0.342                      &0.193                      &0.219                      &0.181                       &0.578                      &0.670                      &0.461                      &0.491                      &0.434                       &0.289                      &0.390                      &0.207                      &0.239                      &0.193                       &0.342                      &0.446                      &0.247                      &0.281                      &0.230                       &0.348                      &0.455                      &0.249                      &0.284                      &0.231                      \\
\multicolumn{1}{c|}{\multirow{-12}{*}{\begin{tabular}[c]{@{}c@{}}\textbf{ID}\\ \textbf{-Rec} \\ (SDSR)\end{tabular}}} & LightSANs                         &0.271                      &0.362                      &0.197                      &0.226                      &0.185                       &0.623                      &0.731                      &0.492                      &0.527                      &0.463                       &0.288                      &0.372                      &0.192                      &0.226                      &0.181                       &0.336                      &0.449                      &0.241                      &0.278                      &0.225                       &0.349                      &0.463                      &0.249                      &0.286                      &0.231                      \\ \hline
\multicolumn{1}{c|}{}                                                                          & BiTGCF                            &0.249                      &0.331                      &0.185                      &0.211                      &0.175                       &0.348                      &0.470                      &0.254                      &0.293                      &0.239                       &0.220                      &0.304                      &0.158                      &0.185                      &0.149                       &0.223                      &0.310                      &0.160                      &0.188                      &0.151                       &0.237                      &0.332                      &0.169                      &0.199                      &0.159                      \\
\multicolumn{1}{c|}{}                                                                          & DTCDR                             &\cellcolor[HTML]{FFCCC9}0.289                      &0.368                      &\cellcolor[HTML]{DAE8FC}0.216                      &\cellcolor[HTML]{FFCCC9}0.245                      &\cellcolor[HTML]{FFCCC9}0.207                       &0.393                      &0.537                      &0.274                      &0.321                      &0.254                       &0.262                      &0.361                      &0.181                      &0.213                      &0.168                       &0.286                      &0.392                      &0.201                      &0.235                      &0.187                       &0.268                      &0.374                      &0.189                      &0.223                      &0.177                      \\
\multicolumn{1}{c|}{}                                                                          & CMF                               &0.202                      &0.282                       &0.147                      &0.173                      &0.140                       &0.312                      &0.417                      &0.229                      &0.263                      &0.216                       &0.255                      &0.343                      &0.184                      &0.212                      &0.173                       &0.240                      &0.329                      &0.172                      &0.201                      &0.161                       &0.243                      &0.345                      &0.171                      &0.204                      &0.161                      \\
\multicolumn{1}{c|}{}                                                                          & CLFM                              &   0.250                   &0.328                      &0.186                      &0.211                      &0.175                       &0.301                      &0.414                      &0.212                      &0.249                      &0.198                       &0.201                      &0.277                      &0.142                      &0.166                      &0.132                       &0.205                      &0.281                      &0.146                      &0.171                      &0.137                       &0.220                      &0.314                      &0.155                      &0.186                      &0.146                      \\
\multicolumn{1}{c|}{}                                                                          & DeepAPF                           &    0.275                  &0.350                      &\cellcolor[HTML]{FFCCC9}0.210                      &0.234                      &0.198                       &0.260                      &0.348                      &0.192                      &0.220                      &0.181                       &0.240                      &0.326                      &0.172                      &0.199                      &0.161                       &0.202                      &0.276                      &0.146                      &0.169                      &0.137                       &0.198                      &0.280                      &0.139                      &0.166                      &0.131                      \\
\multicolumn{1}{c|}{}                                                                          & MoSE                           &    0.236      & 0.344           & 0.170           & 0.205            & 0.182                            &0.351         &  0.599          & 0.235           & 0.315            &  0.251                       &0.288          & 0.405           & 0.202            & 0.240           & 0.207                       &0.280         & 0.419           &0.195            & 0.239            &  0.205                       &0.276          & 0.419           & 0.189           &  0.235           & 0.199                      \\
\multicolumn{1}{c|}{}                                                                          & CGRec                             &    0.258                  &  0.344                    &   0.189                   &   0.217                   &       0.195                &      0.632                &         0.736             &    0.497                  &    0.530                  &    0.476                   &     0.298                 &   0.408                   &      \cellcolor[HTML]{FFCCC9}0.212                &    0.248                  &     0.216                  &  0.351                    &    0.469                  &       \cellcolor[HTML]{DAE8FC}0.253               &   0.291                   &   \cellcolor[HTML]{DAE8FC}0.253                    &     0.354                 &   0.472                   &     0.252                 &      0.290                &      \cellcolor[HTML]{FFCCC9}0.252                \\
\multicolumn{1}{c|}{\multirow{-9}{*}{\begin{tabular}[c]{@{}c@{}}\textbf{ID}\\ \textbf{-Rec} \\ (CDSR)\end{tabular}}}  & SyNCRec                           &   0.228                   &     0.342                 &     0.161                 &    0.198                  &      0.173                 &    0.641                  &   0.766                   &     0.506                 &      0.546                &   0.488                    & \cellcolor[HTML]{FFCCC9}0.301                     &   \cellcolor[HTML]{FFCCC9}0.431                   &             0.211         &      \cellcolor[HTML]{FFCCC9}0.253                &    \cellcolor[HTML]{FFCCC9}0.217                   &            \cellcolor[HTML]{FFCCC9}0.353          &        \cellcolor[HTML]{FFCCC9}0.492             &    \cellcolor[HTML]{FFCCC9}0.250                  &        \cellcolor[HTML]{DAE8FC}0.295              &     \cellcolor[HTML]{FFCCC9}0.252                  &  0.338                    &        0.489              &       0.236               &  0.285                    &    0.241                  \\ \hline
\multicolumn{1}{c|}{}                                                                          & UniSRec                           & 0.181                     &0.257                      &0.129                      &0.153      &0.122                &0.466                       &0.609                      &0.341                      &0.387                      &0.319                      &0.244                       &0.348                        &0.170                      &0.203 &0.159                      &0.274                      &0.390                       &0.195                      &0.232                      &0.184                      &0.253                      &0.363                       &0.175                      &0.211                      &0.165                                                                 \\
\multicolumn{1}{c|}{}                                                                          & E4SRec                            &0.159                      &0.251                      &0.109                      & 0.138                     & 0.131                      &\cellcolor[HTML]{FFCCC9}0.668                      &\cellcolor[HTML]{FFCCC9}0.778                      &0.496                      &0.538                      &0.468                       &0.189                      &0.305                      &0.132                      &0.168                      &0.153                       &0.231                      &0.383                      &0.142                      &0.191                      &0.158                       &0.276                      &0.419                      &0.182                      &0.225                      &0.194                      \\
\multicolumn{1}{c|}{}                                                                          & HLLM                           &0.161                      &0.245                      &0.109                      &0.136                      &0.126                       &0.659                      &0.781                      & 0.515                     &0.555                      &\cellcolor[HTML]{FFCCC9}0.494                       &0.231                      &0.352                      &0.158                      &0.200                      &0.174                       &0.272                      &0.418                      &0.179                      &0.226                      &0.191                       &0.303                      &0.451                      &0.201                      &0.249                      &0.210                      \\
\multicolumn{1}{c|}{}                                                                          & RecFormer                         &    0.136                  &  0.213                    & 0.092                      & 0.116                      & 0.109                       & \cellcolor[HTML]{DAE8FC}0.682                       & \cellcolor[HTML]{DAE8FC}0.793                      & \cellcolor[HTML]{FFCCC9}0.516                      & \cellcolor[HTML]{FFCCC9}0.552                      & 0.485                       & 0.186                      & 0.292                      & 0.128                      & 0.162                      & 0.149                       & 0.207                      & 0.363                      & 0.129                      & 0.179                      &0.151                       & 0.266                      & 0.411                      & 0.173                      & 0.220                      & 0.186                      \\
\multicolumn{1}{c|}{}                                                                          & SAID                              & \multicolumn{1}{l}{0.247} & \multicolumn{1}{l}{0.337} & \multicolumn{1}{l}{0.179} & \multicolumn{1}{l}{0.208} & \multicolumn{1}{l|}{0.186} & \multicolumn{1}{l}{0.573} & \multicolumn{1}{l}{0.687} & \multicolumn{1}{l}{0.435} & \multicolumn{1}{l}{0.472} & \multicolumn{1}{l|}{0.416} & \multicolumn{1}{l}{0.263} & \multicolumn{1}{l}{0.353} & \multicolumn{1}{l}{0.197} & \multicolumn{1}{l}{0.226} & \multicolumn{1}{l|}{0.204} & \multicolumn{1}{l}{0.289} & \multicolumn{1}{l}{0.393} & \multicolumn{1}{l}{0.204} & \multicolumn{1}{l}{0.238} & \multicolumn{1}{l|}{0.208} & \multicolumn{1}{l}{0.295} & \multicolumn{1}{l}{0.413} & \multicolumn{1}{l}{0.200} & \multicolumn{1}{l}{0.238} & \multicolumn{1}{l}{0.201} \\
\multicolumn{1}{c|}{}                                                                          & CALRec                            & \multicolumn{1}{l}{0.154} & \multicolumn{1}{l}{0.245} & \multicolumn{1}{l}{0.102} & \multicolumn{1}{l}{0.131} & \multicolumn{1}{l|}{0.120} & \multicolumn{1}{l}{0.635} & \multicolumn{1}{l}{0.766} & \multicolumn{1}{l}{0.475} & \multicolumn{1}{l}{0.518} & \multicolumn{1}{l|}{0.451} & \multicolumn{1}{l}{0.195} & \multicolumn{1}{l}{0.311} & \multicolumn{1}{l}{0.133} & \multicolumn{1}{l}{0.171} & \multicolumn{1}{l|}{0.154} & \multicolumn{1}{l}{0.224} & \multicolumn{1}{l}{0.373} & \multicolumn{1}{l}{0.139} & \multicolumn{1}{l}{0.187} & \multicolumn{1}{l|}{0.156} & \multicolumn{1}{l}{0.280} & \multicolumn{1}{l}{0.424} & \multicolumn{1}{l}{0.185} & \multicolumn{1}{l}{0.231} & \multicolumn{1}{l}{0.196} \\
\multicolumn{1}{c|}{}                                                                          & ReLLa                             & \multicolumn{1}{l}{0.197} & \multicolumn{1}{l}{0.291} & \multicolumn{1}{l}{0.133} & \multicolumn{1}{l}{0.163} & \multicolumn{1}{l|}{0.145} & \multicolumn{1}{l}{0.603} & \multicolumn{1}{l}{0.737} & \multicolumn{1}{l}{0.412} & \multicolumn{1}{l}{0.486} & \multicolumn{1}{l|}{0.419} & \multicolumn{1}{l}{0.210} & \multicolumn{1}{l}{0.341} & \multicolumn{1}{l}{0.139} & \multicolumn{1}{l}{0.181} & \multicolumn{1}{l|}{0.159} & \multicolumn{1}{l}{0.241} & \multicolumn{1}{l}{0.385} & \multicolumn{1}{l}{0.155} & \multicolumn{1}{l}{0.202} & \multicolumn{1}{l|}{0.171} & \multicolumn{1}{l}{0.292} & \multicolumn{1}{l}{0.438} & \multicolumn{1}{l}{0.191} & \multicolumn{1}{l}{0.239} & \multicolumn{1}{l}{0.201} \\
\multicolumn{1}{c|}{}                                                                          & Lite-LLMRec                       & \multicolumn{1}{l}{0.184} & \multicolumn{1}{l}{0.270} & \multicolumn{1}{l}{0.127} & \multicolumn{1}{l}{0.154} & \multicolumn{1}{l|}{0.141} & \multicolumn{1}{l}{0.661} & \multicolumn{1}{l}{0.776} & \multicolumn{1}{l}{\cellcolor[HTML]{DAE8FC}0.518} & \multicolumn{1}{l}{\cellcolor[HTML]{DAE8FC}0.556} & \multicolumn{1}{l|}{\cellcolor[HTML]{DAE8FC}0.497} & \multicolumn{1}{l}{0.249} & \multicolumn{1}{l}{0.371} & \multicolumn{1}{l}{0.172} & \multicolumn{1}{l}{0.212} & \multicolumn{1}{l|}{0.187} & \multicolumn{1}{l}{0.289} & \multicolumn{1}{l}{0.428} & \multicolumn{1}{l}{0.195} & \multicolumn{1}{l}{0.240} & \multicolumn{1}{l|}{0.206} & \multicolumn{1}{l}{0.318} & \multicolumn{1}{l}{0.463} & \multicolumn{1}{l}{0.213} & \multicolumn{1}{l}{0.260} & \multicolumn{1}{l}{0.220} \\
\multicolumn{1}{c|}{}                                                                          & LLMEmb                          &0.262                      &\cellcolor[HTML]{FFCCC9}0.377                      &0.181                      &0.218                      &0.191                       &0.654                      &0.785                      &0.489                      &0.531                      &0.462                       &0.270                      &0.418                      &0.181                      &0.229                      &0.195                       &0.311                      &0.467                      &0.204                      &0.254                      &0.213                       &0.351                      &\cellcolor[HTML]{FFCCC9}0.502                      &0.240                      &0.289                      &0.245                      \\ \cline{2-27} 
\multicolumn{1}{c|}{\multirow{-9}{*}{\begin{tabular}[c]{@{}c@{}}\textbf{LLM}\\ \textbf{-Rec} \end{tabular}}} & \textbf{Ours}                             &  \cellcolor[HTML]{DAE8FC}\textbf{0.300}         & \cellcolor[HTML]{DAE8FC}\textbf{0.423}           & \textbf{0.206}           & \cellcolor[HTML]{DAE8FC}\textbf{0.246}             & \cellcolor[HTML]{DAE8FC}\textbf{0.213}   & \textbf{0.609}          & \textbf{0.751}           & \textbf{0.458}           & \textbf{0.505}            &  \textbf{0.440}    & \cellcolor[HTML]{DAE8FC}\textbf{0.363}          & \cellcolor[HTML]{DAE8FC}\textbf{0.504}            & \cellcolor[HTML]{DAE8FC}\textbf{0.258}             & \cellcolor[HTML]{DAE8FC}\textbf{0.303}           & \cellcolor[HTML]{DAE8FC}\textbf{0.263}                  &  \cellcolor[HTML]{DAE8FC}\textbf{0.359}           &  \cellcolor[HTML]{DAE8FC}\textbf{0.513}     &  \textbf{0.244}         & \cellcolor[HTML]{FFCCC9}\textbf{0.293}           & \textbf{0.247}           & \cellcolor[HTML]{DAE8FC}\textbf{0.400}            & \cellcolor[HTML]{DAE8FC}\textbf{0.550}     & \cellcolor[HTML]{DAE8FC}\textbf{0.285}          & \cellcolor[HTML]{DAE8FC}\textbf{0.333}           & \cellcolor[HTML]{DAE8FC}\textbf{0.286}                      \\ 
\bottomrule
\end{tabular}
\vspace{5pt} 
\end{table*}

\subsection{Experimental Setup} \label{subsection:setup}
\subsubsection{Baselines}
We assessed the performance of our model by comparing it to three categories of baselines: (1) IDRec for Single-Domain Sequential Recommendation (\textbf{IDRec-SDSR}), (2) IDRec for Cross-Domain (Sequential) Recommendation (\textbf{IDRec-CDSR}), and (3) LLMRec for Cross-Domain Sequential Recommendation (\textbf{LLMRec}) as shown in Table \ref{tab:result_1}. 
A detailed description of baselines can be found in Section \ref{section:related_work} and Appendix \ref{section:baselines}.

\subsubsection{Evaluation Settings}
We used the \textit{leave-one-out} method to evaluate the recommendation performance.
For each user interaction sequence, the sequence was split into three segments: the final item was used as test data, the second-to-last item served as validation data, and the remaining preceding items were utilized as training data. 
The evaluation was performed separately for each domain, determined by the domain of the last item in the test data.

Considering the extensive number of items, it was infeasible to use all items as test candidates due to computational limitations. 
To address this, we used a common approach that pairs the ground-truth item (positive sample) with 99 randomly selected items (negative samples) with which the user has no prior interaction history. 
We focused on standard evaluation metrics, including \textit{Hit Ratio} (\textbf{HR}), \textit{Normalized Discounted Cumulative Gain} (\textbf{NDCG}), and \textit{Mean Reciprocal Rank} (\textbf{MRR}).

\subsection{Implementation Details} \label{subsection: hyperparameter_setting}

\noindent\textbf{(1) LLM Backbone}
We employed \texttt{TinyLlama-1.1B}~\cite{zhang2024tinyllama} as the LLM backbone to meet the strict efficiency requirements of large-scale real-world recommendation serving.
In this LLM, the hidden size and the context length were set to 2048 and 4096, respectively.
The number of heads was 32, and the number of transformer layers was 22.
The vocabulary size in the tokenizer was set to 32000.

\noindent\textbf{(2) Hyperparameter Setting}
In the \textit{Domain-Gated Dual Sequential Encoder}, the hidden size of the transformer, the number of attention heads and the number of transformer layers were set to 2048, 16 and 2 respectively.
The hyperparameter $p$ in the \textit{Dual-Sampling Token-to-Item Contrastive Learning} was set to 0.4.
The harmonic factor $\eta$ was set to 0.5.

\noindent\textbf{(3) Model Training}
The AdamW optimizer was used to update all parameters. 
The batch size and the number of training epochs were 4 and 25, respectively. 
To ensure a fair comparison, all LLMRec baselines were implemented with the same \texttt{TinyLlama-1.1B} backbone as our model, and the context length, batch size, and optimizer were matched across all baseline models.
All experiments were conducted on four NVIDIA A100 40GB GPUs.

\begin{table*}[t] \footnotesize
\centering
\caption{Model performance comparison on \textit{\textbf{Telco}}, with the top two methods highlighted in \setlength{\fboxsep}{0pt}\colorbox[HTML]{DAE8FC}{blue} and \setlength{\fboxsep}{0pt}\colorbox[HTML]{FFCCC9}{red}, respectively.
}
    \renewcommand{\arraystretch}{1.05}
\label{tab:result_2}
\setlength{\tabcolsep}{0.79pt}
\begin{tabular}{cc|ccccc|ccccc|ccccc|ccccc|ccccc}
\toprule
\multicolumn{2}{c|}{\cellcolor[HTML]{EFEFEF}\textbf{Domain}}                                                                                & \multicolumn{5}{c|}{\cellcolor[HTML]{EFEFEF}\textbf{Web-View}}                                                       & \multicolumn{5}{c|}{\cellcolor[HTML]{EFEFEF}\textbf{Call-Log}}                                                       & \multicolumn{5}{c|}{\cellcolor[HTML]{EFEFEF}\textbf{PoI}}                                                    & \multicolumn{5}{c|}{\cellcolor[HTML]{EFEFEF}\textbf{Benefits}}                                                      & \multicolumn{5}{c}{\cellcolor[HTML]{EFEFEF}\textbf{Shopping}}                                                        \\ \hline\hline
\multicolumn{1}{c|}{}                                                                          &                                   & \multicolumn{2}{c}{HR}                      & \multicolumn{2}{c}{NDCG}                    & MRR                   & \multicolumn{2}{c}{HR}                      & \multicolumn{2}{c}{NDCG}                    & MRR                   & \multicolumn{2}{c}{HR}                      & \multicolumn{2}{c}{NDCG}                    & MRR                   & \multicolumn{2}{c}{HR}                      & \multicolumn{2}{c}{NDCG}                    & MRR                   & \multicolumn{2}{c}{HR}                      & \multicolumn{2}{c}{NDCG}                    & MRR                  \\ \cline{3-27} 
\multicolumn{1}{c|}{\multirow{-2}{*}{\textbf{Types}}}                                                   & \multirow{-2}{*}{\textbf{Models}} & @5                   & @10                  & @5                   & @10                  & @10                   & @5                   & @10                  & @5                   & @10                  & @10                   & @5                   & @10                  & @5                   & @10                  & @10                   & @5                   & @10                  & @5                   & @10                  & @10                   & @5                   & @10                  & @5                   & @10                  & @10                  \\ \hline
\multicolumn{1}{c|}{}                                                                          & GRU4Rec                           & 0.946                & 0.976                & 0.842                & 0.852                & 0.811                 & 0.700                & 0.845                & 0.507                & 0.554                & 0.462                 & 0.906                & 0.969                & 0.746                & 0.766                & 0.701                 & 0.874                & 0.945                & 0.680                & 0.703                & 0.625                 & 0.328                & 0.486                & 0.227                & 0.278                & 0.214                \\
\multicolumn{1}{c|}{}                                                                          & SASRec                            & 0.944                & 0.975                & 0.843                & 0.853                & 0.812                 & 0.707                & 0.846                & 0.532                & 0.577                & 0.493                 &0.893                 &0.965                &0.722                &0.746                &0.675                 & 0.892                & 0.955                & 0.705                & 0.725                & 0.650                & 0.391               & 0.541                & 0.290                & 0.338                & 0.277                \\
\multicolumn{1}{c|}{}                                                                          & BERT4Rec                          & 0.936                     &0.972                      &0.821                      &0.833                      &0.787                       &\cellcolor[HTML]{FFCCC9}0.716                      &\cellcolor[HTML]{FFCCC9}0.857                      &0.529                      &0.575                      &0.486                       &0.901                      &0.962                      &0.742                      &0.761                      &0.696                       &0.884                      &0.955                      &0.692                      &0.716                      &0.638                       &\cellcolor[HTML]{FFCCC9}0.411                      &0.554                      &0.293                      &0.340                      &0.274                      \\
\multicolumn{1}{c|}{}                                                                          & S3Rec                             &0.833                      &0.913                      &0.677                      &0.703                      &0.636                       &0.714                      &0.849                      &\cellcolor[HTML]{FFCCC9}0.546                      &\cellcolor[HTML]{FFCCC9}0.590                      &\cellcolor[HTML]{FFCCC9}0.509                       &0.624                      &0.654                      &0.515                      &0.525                      &0.483                       &0.884                      &0.949                      &0.691                      &0.712                      &0.635                       &0.406                      &\cellcolor[HTML]{FFCCC9}0.557                      &\cellcolor[HTML]{FFCCC9}0.299                      &\cellcolor[HTML]{FFCCC9}0.347                      &0.284                      \\
\multicolumn{1}{c|}{}                                                                          & NextItnet                         &0.939                      &0.973                      &0.827                      &0.839                      &0.794                       &0.649                      &0.810                      &0.454                      &0.507                      &0.412                       &0.903                      &\cellcolor[HTML]{FFCCC9}0.970                      &0.743                      &0.765                      &0.699                       &0.865                      &0.937                      &0.641                      &0.664                      &0.575                       &0.326                      &0.486                      &0.234                      &0.285                      &0.225                      \\
\multicolumn{1}{c|}{}                                                                          & SINE                              &0.931                      &0.968                      &0.817                      &0.829                      &0.784                       &0.602                      &0.781                      &0.400                      &0.458                      &0.357                       &0.830                      &0.939                      &0.614                      &0.650                      &0.558                       &0.859                      &0.918                      &0.635                      &0.654                      &0.567                       &0.104                      &0.314                      &0.050                      &0.117                      &0.060                      \\
\multicolumn{1}{c|}{}                                                                          & STAMP                             &\cellcolor[HTML]{FFCCC9}0.949                      &\cellcolor[HTML]{DAE8FC}0.977                      &0.852                      &0.862                      &0.823                       &0.692                      &0.839                      &0.505                      &0.552                      &0.462                       &0.899                      &0.966                      &0.734                      &0.756                      &0.688                       &0.884                      &0.950                      &0.698                      &0.720                      &0.645                       &0.341                      &0.489                      &0.245                      &0.293                      &0.233                      \\
\multicolumn{1}{c|}{}                                                                          & TransRec                          &0.933                      &0.969                      &0.822                      &0.834                      &0.789                       &0.637                      &0.803                      &0.468                      &0.522                      &0.434                       &0.873                      &0.958                      &0.694                     &0.722                      &0.645                       &0.858                      &0.920                      &0.666                      &0.686                      &0.610                       &0.315                      &0.427                      &0.231                      &0.267                      &0.218                      \\
\multicolumn{1}{c|}{\multirow{-12}{*}{\begin{tabular}[c]{@{}c@{}}\textbf{ID}\\ \textbf{-Rec} \\ (SDSR)\end{tabular}}} & LightSANs                         &0.938                      &0.972                      &0.837                      &0.848                      &0.808                       &0.702                      &0.841                      &0.531                      &0.576                      &0.493                       &0.902                      &0.968                      &0.742                      &0.763                      &0.687                       &0.890                      &\cellcolor[HTML]{FFCCC9}0.955                      &0.701                      &0.723                      &0.647                       &0.397                      &0.529                      &0.292                      &0.335                      &0.275                      \\ \hline
\multicolumn{1}{c|}{}                                                                          & BiTGCF                            &0.931                      &0.967                      &0.823                      &0.835                      &0.791                       &0.644                     &0.782                      &0.483                      &0.527                      &0.448                       &0.625                      &0.762                      &0.475                      &0.520                      &0.444                       &0.854                      &0.915                      &0.725                      &0.745                      &\cellcolor[HTML]{DAE8FC}0.690                       &0.378                      &0.498                      &0.292                      &0.330                      &0.279                      \\
\multicolumn{1}{c|}{}                                                                          & DTCDR                             &0.944                      &0.974                      &0.842                      &0.853                      &0.812                       &0.652                      &0.788                      &0.493                      &0.538                      &0.459                       &0.634                      &0.776                      &0.490                      &0.536                      &0.462                       &0.826                      &0.923                      &0.621                      &0.653                      &0.566                       &0.293                      &0.456                      &0.192                      &0.244                      &0.180                      \\
\multicolumn{1}{c|}{}                                                                          & CMF                               &0.946                      &0.975                      &0.848                      &0.858                      &0.819                       &0.685                      &0.808                      &0.534                      &0.574                      &0.501                       &0.668                      &0.790                      &0.533                      &0.578                      &0.512                       &0.717                      &0.775                      &0.658                      &0.677                      &0.646                       &0.325                      &0.454                      &0.250                      &0.292                      &0.243                      \\
\multicolumn{1}{c|}{}                                                                          & CLFM                              &0.949                      &0.976                      &\cellcolor[HTML]{FFCCC9}0.853                      &\cellcolor[HTML]{FFCCC9}0.862                      &0.824                       &0.679                      &0.802                      &0.520                      &0.560                      &0.484                       &0.645                      &0.769                      &0.511                      &0.551                      &0.483                       &0.823                      &0.876                      &0.727                      &0.744                      &0.702                       &0.305                      &0.422                      &0.229                      &0.267                      &0.219                      \\
\multicolumn{1}{c|}{}                                                                          & DeepAPF                           &0.949                      &0.974                      &0.858                      &0.867                      &\cellcolor[HTML]{FFCCC9}0.831                       &0.653                      &0.778                      &0.521                      &0.561                      &0.494                       &0.657                      &0.775                      &0.544                      &0.582                      &0.529                       &0.828                      &0.892                      &\cellcolor[HTML]{DAE8FC}0.745                      &\cellcolor[HTML]{DAE8FC}0.766                      &0.726                       &0.307                      &0.422                      &0.242                      &0.279                      &0.236                      \\
\multicolumn{1}{c|}{}                                                                          & MoSE                           &    0.887          &  0.927          & 0.817           & 0.830            & 0.803      &  0.374         & 0.405           & 0.343            &0.354            & 0.351      & 0.357          & 0.373           & 0.334            & 0.339           & 0.347     & 0.245          & 0.327           & 0.202           & 0.228            & 0.230       & 0.292          & 0.333           &  0.231          &  0.244           & 0.235                      \\
\multicolumn{1}{c|}{}                                                                          & CGRec                             &    0.919                  &  0.959                    &   0.820                   &   0.833                   &       0.794                &      0.692                &         0.832             &    0.529                  &   0.575                  &    0.503                   &     0.885                 &   0.958                   &      0.738                &    0.763                  &    0.702                  &  \cellcolor[HTML]{FFCCC9}0.901                    &   0.943                  &      0.731               &   0.744                   &   0.682                    &     0.371                 &   0.506                   &     0.281                 &      0.324                &      \cellcolor[HTML]{FFCCC9}0.293                \\
\multicolumn{1}{c|}{\multirow{-9}{*}{\begin{tabular}[c]{@{}c@{}}\textbf{ID}\\ \textbf{-Rec} \\ (CDSR)\end{tabular}}}  & SyNCRec                           &   0.942                   &     0.973                 &     \cellcolor[HTML]{DAE8FC}0.857                 &    \cellcolor[HTML]{DAE8FC}0.867                  &      \cellcolor[HTML]{DAE8FC}0.834                 &    0.695                  &   0.828                   &     0.529                 &      0.572                &   0.501                    & \cellcolor[HTML]{FFCCC9}0.906                     &   0.969                   &             \cellcolor[HTML]{DAE8FC}0.764         &      \cellcolor[HTML]{DAE8FC}0.784                &    \cellcolor[HTML]{DAE8FC}0.727                   &            0.883          &        0.946             &    0.727                  &        0.747              &     \cellcolor[HTML]{FFCCC9}0.686                  &  0.325                    &        0.480              &       0.233               &  0.282                    &    0.246                  \\ \hline
\multicolumn{1}{c|}{}                                                                          & UniSRec                           & 0.830                     &0.901                      &0.686                      &0.709      &0.647                &0.618                       &0.763                      &0.429                      &0.476                      &0.386                      &0.672                       &0.828                        &0.503                      &0.554 &0.469                      &0.793                      &0.881                       &0.574                      &0.604                      &0.514                      &0.205                      &0.412                       &0.107                      &0.174                      &0.103                                                                 \\
\multicolumn{1}{c|}{}                                                                          & E4SRec                            &0.917                      &0.959                      &0.807                      &0.822                      &0.772                       &0.568                      &0.739                      &0.379                      &0.441                      &0.358                       &0.821                      &0.910                      &0.651                      &0.676                      &0.601                       &0.843                      &0.917                      &0.655                      &0.671                      &0.599                       &0.270                      &0.419                      &0.173                      &0.224                      &0.194                      \\
\multicolumn{1}{c|}{}                                                                          & HLLM                           &0.916                      &0.958                      &0.803                      &0.817                      &0.773                       &0.599                      &0.775                      &0.410                      &0.468                      &0.384                       &0.851                      &0.938                      &0.681                      &0.709                      & 0.640 & 0.866                       &0.932                      &0.702                      &0.723                      &0.659                      &0.277                       &0.444                      &0.186                      &0.240                      &0.202                                           \\
\multicolumn{1}{c|}{}                                                                          & RecFormer                         &  0.910                    &0.953                      &0.796                      &0.810                      &0.766                       &0.581                      &0.751                      &0.397                      &0.453                      &0.372                       &0.835                      &0.927                      &0.663                      &0.693                      &0.622                       &0.866                      &0.937                      &0.670                      &0.693                      &0.617                       &0.261                      &0.415                      &0.172                      &0.221                      &0.187                      \\
\multicolumn{1}{c|}{}                                                                          & SAID                              & \multicolumn{1}{l}{0.931} & \multicolumn{1}{l}{0.963} & \multicolumn{1}{l}{0.837} & \multicolumn{1}{l}{0.847} & \multicolumn{1}{l|}{0.811} & \multicolumn{1}{l}{0.661} & \multicolumn{1}{l}{0.794} & \multicolumn{1}{l}{0.511} & \multicolumn{1}{l}{0.554} & \multicolumn{1}{l|}{0.490} & \multicolumn{1}{l}{0.849} & \multicolumn{1}{l}{0.936} & \multicolumn{1}{l}{0.699} & \multicolumn{1}{l}{0.728} & \multicolumn{1}{l|}{0.665} & \multicolumn{1}{l}{0.857} & \multicolumn{1}{l}{0.918} & \multicolumn{1}{l}{0.697} & \multicolumn{1}{l}{0.716} & \multicolumn{1}{l|}{0.655} & \multicolumn{1}{l}{0.237} & \multicolumn{1}{l}{0.364} & \multicolumn{1}{l}{0.184} & \multicolumn{1}{l}{0.225} & \multicolumn{1}{l}{0.211} \\
\multicolumn{1}{c|}{}                                                                          & CALRec                            & \multicolumn{1}{l}{0.905} & \multicolumn{1}{l}{0.948} & \multicolumn{1}{l}{0.782} & \multicolumn{1}{l}{0.796} & \multicolumn{1}{l|}{0.748} & \multicolumn{1}{l}{0.360} & \multicolumn{1}{l}{0.561} & \multicolumn{1}{l}{0.223} & \multicolumn{1}{l}{0.288} & \multicolumn{1}{l|}{0.227} & \multicolumn{1}{l}{0.716} & \multicolumn{1}{l}{0.894} & \multicolumn{1}{l}{0.490} & \multicolumn{1}{l}{0.548} & \multicolumn{1}{l|}{0.446} & \multicolumn{1}{l}{0.768} & \multicolumn{1}{l}{0.887} & \multicolumn{1}{l}{0.550} & \multicolumn{1}{l}{0.590} & \multicolumn{1}{l|}{0.501} & \multicolumn{1}{l}{0.171} & \multicolumn{1}{l}{0.273} & \multicolumn{1}{l}{0.112} & \multicolumn{1}{l}{0.144} & \multicolumn{1}{l}{0.130} \\
\multicolumn{1}{c|}{}                                                                          & ReLLa                             & \multicolumn{1}{l}{0.896} & \multicolumn{1}{l}{0.941} & \multicolumn{1}{l}{0.764} & \multicolumn{1}{l}{0.779} & \multicolumn{1}{l|}{0.728} & \multicolumn{1}{l}{0.356} & \multicolumn{1}{l}{0.571} & \multicolumn{1}{l}{0.203} & \multicolumn{1}{l}{0.273} & \multicolumn{1}{l|}{0.203} & \multicolumn{1}{l}{0.790} & \multicolumn{1}{l}{0.913} & \multicolumn{1}{l}{0.521} & \multicolumn{1}{l}{0.568} & \multicolumn{1}{l|}{0.464} & \multicolumn{1}{l}{0.761} & \multicolumn{1}{l}{0.888} & \multicolumn{1}{l}{0.509} & \multicolumn{1}{l}{0.551} & \multicolumn{1}{l|}{0.450} & \multicolumn{1}{l}{0.183} & \multicolumn{1}{l}{0.313} & \multicolumn{1}{l}{0.113} & \multicolumn{1}{l}{0.156} & \multicolumn{1}{l}{0.132} \\
\multicolumn{1}{c|}{}                                                                          & Lite-LLMRec                       & \multicolumn{1}{l}{0.919} & \multicolumn{1}{l}{0.960} & \multicolumn{1}{l}{0.809} & \multicolumn{1}{l}{0.823} & \multicolumn{1}{l|}{0.780} & \multicolumn{1}{l}{0.603} & \multicolumn{1}{l}{0.777} & \multicolumn{1}{l}{0.417} & \multicolumn{1}{l}{0.474} & \multicolumn{1}{l|}{0.391} & \multicolumn{1}{l}{0.848} & \multicolumn{1}{l}{0.934} & \multicolumn{1}{l}{0.679} & \multicolumn{1}{l}{0.707} & \multicolumn{1}{l|}{0.638} & \multicolumn{1}{l}{0.871} & \multicolumn{1}{l}{0.943} & \multicolumn{1}{l}{0.697} & \multicolumn{1}{l}{0.720} & \multicolumn{1}{l|}{0.652} & \multicolumn{1}{l}{0.283} & \multicolumn{1}{l}{0.418} & \multicolumn{1}{l}{0.188} & \multicolumn{1}{l}{0.232} & \multicolumn{1}{l}{0.201} \\
\multicolumn{1}{c|}{}                                                                          & LLMEmb                          &0.926                      &0.961                      &0.807                      &0.818                      &0.773                       &0.491                      &0.652                      &0.355                      &0.407                      &0.348                       &0.851                      &0.939                      &0.653                      &0.682                      &0.602                       &0.823                      &0.917                      &0.628                      &0.659                      &0.581                       &0.315                      &0.436                      &0.215                      &0.254                      &0.217                      \\ \cline{2-27} 
\multicolumn{1}{c|}{\multirow{-9}{*}{\begin{tabular}[c]{@{}c@{}}\textbf{LLM}\\ \textbf{-Rec} \end{tabular}}} & \textbf{Ours}                             &  \cellcolor[HTML]{DAE8FC}\textbf{0.950}         & \cellcolor[HTML]{FFCCC9}\textbf{0.976}           & \textbf{0.849}           & \textbf{0.858}             & \textbf{0.820}   & \cellcolor[HTML]{DAE8FC}\textbf{0.747}          & \cellcolor[HTML]{DAE8FC}\textbf{0.872}           & \cellcolor[HTML]{DAE8FC}\textbf{0.570}           & \cellcolor[HTML]{DAE8FC}\textbf{0.611}            &  \cellcolor[HTML]{DAE8FC}\textbf{0.535}    & \cellcolor[HTML]{DAE8FC}\textbf{0.916}          & \cellcolor[HTML]{DAE8FC}\textbf{0.970}            & \cellcolor[HTML]{FFCCC9}\textbf{0.754}             & \cellcolor[HTML]{FFCCC9}\textbf{0.772}           & \cellcolor[HTML]{FFCCC9}\textbf{0.708}                  &  \cellcolor[HTML]{DAE8FC}\textbf{0.913}           &  \cellcolor[HTML]{DAE8FC}\textbf{0.964}     &  \cellcolor[HTML]{FFCCC9}\textbf{0.733}         & \cellcolor[HTML]{FFCCC9}\textbf{0.750}           & \textbf{0.681}           & \cellcolor[HTML]{DAE8FC}\textbf{0.423}            & \cellcolor[HTML]{DAE8FC}\textbf{0.588}     & \cellcolor[HTML]{DAE8FC}\textbf{0.300}          & \cellcolor[HTML]{DAE8FC}\textbf{0.353}           & \cellcolor[HTML]{DAE8FC}\textbf{0.300}                      \\ 
\bottomrule
\end{tabular}
\vspace{5pt} 
\end{table*}

\subsection{Performance Evaluation (RQ1)}  \label{subsection:performance_evaluation}
In Table \ref{tab:result_1} (\textbf{Amazon}) and Table \ref{tab:result_2} (\textbf{Telco}), we reported the performance of all the baselines and our model.\footnote{Pairwise CDR/CDSR baselines (BiTGCF, CMF, DTCDR, DeepAPF) are trained on each two-domain pair and the average performance across pairs is reported per target domain; full per-pair results are provided in Appendix~\ref{section: pairwise}.}

\noindent \textbf{Our model showed overall improved recommendation performance for most domains in both datasets compared to the state-of-the-art IDRec and LLMRec models}.
Our model demonstrated significant improvements over the best-performing baseline (second-best for each domain) in HR@5 on the Amazon dataset, achieving gains of \setlength{\fboxsep}{0pt}\colorbox[HTML]{DAE8FC}{\textbf{+3.81\%}} (\textit{Books}), \setlength{\fboxsep}{0pt}\colorbox[HTML]{DAE8FC}{\textbf{+20.6\%}} (\textit{Jewelry}), \setlength{\fboxsep}{0pt}\colorbox[HTML]{DAE8FC}{\textbf{+1.70\%}} (\textit{Outdoors}), and \setlength{\fboxsep}{0pt}\colorbox[HTML]{DAE8FC}{\textbf{+10.8\%}} (\textit{Toys}). 
Similarly, on the Telco dataset, it outperformed with increases of \setlength{\fboxsep}{0pt}\colorbox[HTML]{DAE8FC}{\textbf{+0.11\%}} (\textit{Web}-\textit{View}), \setlength{\fboxsep}{0pt}\colorbox[HTML]{DAE8FC}{\textbf{+4.33\%}} (\textit{Call}-\textit{Log}), \setlength{\fboxsep}{0pt}\colorbox[HTML]{DAE8FC}{\textbf{+1.10\%}} (\textit{PoI}), \setlength{\fboxsep}{0pt}\colorbox[HTML]{DAE8FC}{\textbf{+1.33\%}} (\textit{Benefits}-\textit{Use}), and \setlength{\fboxsep}{0pt}\colorbox[HTML]{DAE8FC}{\textbf{+2.92\%}} (\textit{Shopping}). 
In the \textit{Games} domain of the Amazon dataset, although our model did not match the best performing baseline (RecFormer), it showed consistently superior performance across most domains.
In contrast, other baseline models such as RecFormer, E4SRec, DTCDR and DeepAPF excelled in certain domains, but showed significantly lower performance in others.

\subsection{Discussion of the Negative Transfer (RQ2)} \label{subsection:diss_nt}

\begin{figure}[htbp]
\centering
\includegraphics[width=0.94\linewidth]{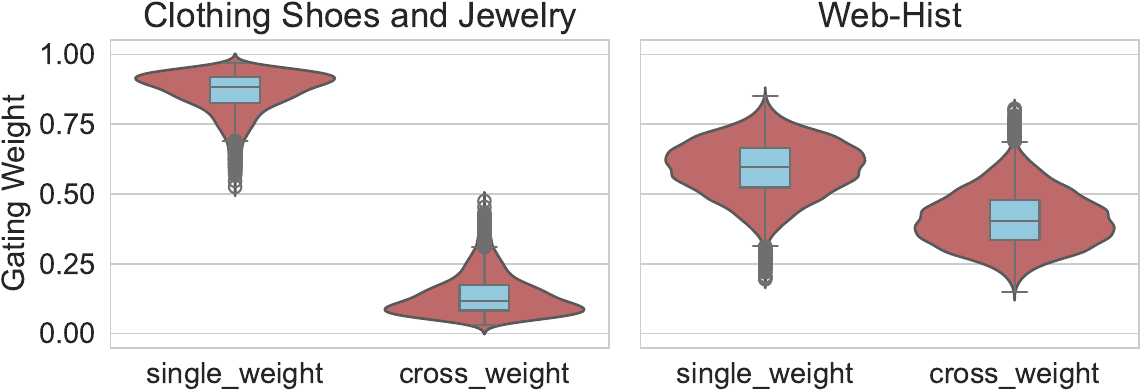}
\caption{Visualization of Single- and Cross-Expert Weights}
\label{fig:gating_weight}
\end{figure}

The negative transfer issue becomes evident when comparing $\circled{1}$ the performance of a domain-specific SDSR model trained solely on single-domain sequences with $\circled{2}$ that of a CDSR model trained on cross-domain sequences (i.e., $\circled{2}$-$\circled{1}$). 
For the domain-specific SDSR model, we utilized SASRec~\cite{kang2018self}, a state-of-the-art ID-based model designed for training on single-domain sequences.

As presented in Table \ref{tab:result_3}, all IDRec and LLMRec models, except ours, underperformed the SDSR approach in most domains in both datasets.
In addition, the LLMRec models, with the exception of ours, showed more pronounced negative transfer effects than the IDRec models. 
\textbf{However, our model showed mitigated negative transfer effects compared to both the IDRec and LLMRec approaches in all domains, with particularly better results in the \textit{Jewelry}, \textit{Call}-\textit{Log} and \textit{Shopping} domains where other models struggled with negative transfer.}
As shown in Table \ref{tab:result_3}, in the \textit{Jewelry} and \textit{Call}-\textit{Log} domains, six baselines (excluding ours) exhibited the performance degradation in HR@5, with five baselines showing similar issues in the \textit{Shopping} domain.
Fig.~\ref{fig:gating_weight} shows how the trained model allocates gating weights between single- and cross-domain experts when predicting items from two representative domains in the test dataset.
In \textit{Jewelry}, where negative transfer is prominent, the model strongly favors the single-domain expert. In contrast, in \textit{Web-Hist}, it assigns substantial weight to the cross-domain expert, suggesting effective use of cross-domain knowledge.

\begin{table}[htbp] \footnotesize
\centering
\caption{The performance gain(\%) of CDSR baselines on cross-domain sequences over SASRec~\cite{kang2018self} on single-domain sequences is shown, with negative in \setlength{\fboxsep}{0pt}\colorbox[HTML]{FFCCC9}{red}.}
    \renewcommand{\arraystretch}{1.2}
\label{tab:result_3}
\setlength{\tabcolsep}{0.8pt}
\begin{tabular}{c|c|c|ccccc|ccccc}
\toprule
\rowcolor[HTML]{EFEFEF} 
\cellcolor[HTML]{EFEFEF}                                                       & \cellcolor[HTML]{EFEFEF}                                  & \textbf{Dataset} & \multicolumn{5}{c|}{\cellcolor[HTML]{EFEFEF}\textbf{Amazon}}                                                       & \multicolumn{5}{c}{\cellcolor[HTML]{EFEFEF}\textbf{Telco}}                                         \\ \cline{3-13} 
\rowcolor[HTML]{EFEFEF} 
\multirow{-2}{*}{\cellcolor[HTML]{EFEFEF}\textbf{Type}}                        & \multirow{-2}{*}{\cellcolor[HTML]{EFEFEF}\textbf{Model}} & \textbf{Domains} & \textbf{B}                    & \textbf{G}               & \textbf{J}                    & \textbf{O} & \textbf{T} & \textbf{W}                    & \textbf{C} & \textbf{P} & \textbf{B}                    & \textbf{S} \\ \hline
                                                                               &                                                           & HR@5             & \cellcolor[HTML]{FFCCC9}-12.7 & \cellcolor[HTML]{FFCCC9}-35.7                    & \cellcolor[HTML]{FFCCC9}\cellcolor[HTML]{FFCCC9}-45.1                         & \cellcolor[HTML]{FFCCC9}-40.5      & \cellcolor[HTML]{FFCCC9}-39.0      & \cellcolor[HTML]{ECF4FF}\cellcolor[HTML]{FFCCC9}-10.6 & \cellcolor[HTML]{FFCCC9}-30.6      & \cellcolor[HTML]{FFCCC9}-23.8      & \cellcolor[HTML]{FFCCC9}-23.0                         & \cellcolor[HTML]{FFCCC9}-38.0      \\
                                                                               &                                                           & NDCG@5           & +7.7                         & \cellcolor[HTML]{FFCCC9}-15.7                    & \cellcolor[HTML]{FFCCC9}-28.6 & \cellcolor[HTML]{FFCCC9}-22.4      & \cellcolor[HTML]{FFCCC9}-19.0      & \cellcolor[HTML]{FFCCC9}-2.70                         & \cellcolor[HTML]{FFCCC9}-20.9      & \cellcolor[HTML]{FFCCC9}-6.40      & \cellcolor[HTML]{FFCCC9}-3.30 & \cellcolor[HTML]{FFCCC9}-12.7      \\
                                                                               & \multirow{-3}{*}{CLFM}                                    & MRR@10              & \cellcolor[HTML]{FFCCC9}-1.2                         &+14.1                          &+20.1                               &+12.2            &+21.8            &+17.1                               &+30.3            &+46.1            &+60.7                               &+54.7            \\ \cline{2-13} 
                                                                               &                                                           & HR@5             &+2.5                               &+47.6                          &\cellcolor[HTML]{FFCCC9}-10.0                               &+3.90            &+13.6            &\cellcolor[HTML]{FFCCC9}-2.90                               &\cellcolor[HTML]{FFCCC9}-6.30            &+31.3            &+10.0                               &+0.50            \\
                                                                               &                                                           & NDCG@5           & \cellcolor[HTML]{FFCCC9}-1.9                              & +64.3 &\cellcolor[HTML]{FFCCC9}-10.1                               &+4.50            &+14.6            &\cellcolor[HTML]{FFCCC9}-2.4                               &\cellcolor[HTML]{FFCCC9}-12.2            &+43.3            &+22.9                               &+15.4            \\
                                                                               & \multirow{-3}{*}{CGRec}                                   & MRR@10              &\cellcolor[HTML]{FFCCC9}-4.3                               &+70.1                          &\cellcolor[HTML]{FFCCC9}-1.3                               &+12.2            &+21.9            &\cellcolor[HTML]{FFCCC9}-1.7                               &\cellcolor[HTML]{FFCCC9}-12.0            &+46.1            &+27.8                               &+30.5            \\ \cline{2-13} 
                                                                               &                                                           & HR@5             &\cellcolor[HTML]{FFCCC9}-9.4                               &+49.7                          &\cellcolor[HTML]{FFCCC9}-9.20                               &+4.30            &+8.40            &\cellcolor[HTML]{FFCCC9}-0.50                               &\cellcolor[HTML]{FFCCC9}-6.00            &+34.5            &+7.80                               &\cellcolor[HTML]{FFCCC9}-11.9            \\
                                                                               &                                                           & NDCG@5           &  \cellcolor[HTML]{FFCCC9}-16.3                             &+67.3                          &\cellcolor[HTML]{FFCCC9}-10.6                               &+3.60            &+7.40            &+2.00                               &\cellcolor[HTML]{FFCCC9}-12.3            &+48.2            &+22.2                               &\cellcolor[HTML]{FFCCC9}-4.50            \\
\multirow{-9}{*}{\textbf{\begin{tabular}[c]{@{}c@{}}ID\\ -Rec\end{tabular}}}   & \multirow{-3}{*}{SyNCRec}                                 & MRR@10              &\cellcolor[HTML]{FFCCC9}-15.2                               &+74.2                          &\cellcolor[HTML]{FFCCC9}-1.00                               &+11.9            &+16.7            &+3.20                               &\cellcolor[HTML]{FFCCC9}-12.3            &+51.2            &+28.5                               &+9.50            \\ \hline
                                                                               &                                                           & HR@5             & \cellcolor[HTML]{FFCCC9}-2.1                              &+33.8                          &\cellcolor[HTML]{FFCCC9}-20.6                               &\cellcolor[HTML]{FFCCC9}-14.4            &\cellcolor[HTML]{FFCCC9}-5.10            &\cellcolor[HTML]{FFCCC9}-1.70                               &\cellcolor[HTML]{FFCCC9}-10.5            &+26.0            &+4.60                               &\cellcolor[HTML]{FFCCC9}-35.9            \\
                                                                               &                                                           & NDCG@5           & \cellcolor[HTML]{FFCCC9}-7.2                              &+43.9                          &\cellcolor[HTML]{FFCCC9}-16.6                               &\cellcolor[HTML]{FFCCC9}-15.4            &\cellcolor[HTML]{FFCCC9}-9.10            &\cellcolor[HTML]{FFCCC9}-0.30                               &\cellcolor[HTML]{FFCCC9}-15.2            &+35.7            &+17.2                               &\cellcolor[HTML]{FFCCC9}-24.3            \\
                                                                               & \multirow{-3}{*}{SAID}                                    & MRR@10              &\cellcolor[HTML]{FFCCC9}-9.1                               &+48.5                          &\cellcolor[HTML]{FFCCC9}-6.80                               &\cellcolor[HTML]{FFCCC9}-7.90            &\cellcolor[HTML]{FFCCC9}-2.50            &+0.40                               &\cellcolor[HTML]{FFCCC9}-14.2            &+38.4            &+22.8                               &\cellcolor[HTML]{FFCCC9}-6.00            \\ \cline{2-13} 
                                                                               &                                                           & HR@5             &  \cellcolor[HTML]{FFCCC9}-21.7                             &+40.9                          &\cellcolor[HTML]{FFCCC9}-36.6                               &\cellcolor[HTML]{FFCCC9}-28.7            &\cellcolor[HTML]{FFCCC9}-6.10            &\cellcolor[HTML]{FFCCC9}-5.40                               &\cellcolor[HTML]{FFCCC9}-51.8            &+17.2            &\cellcolor[HTML]{FFCCC9}-7.10                               &\cellcolor[HTML]{FFCCC9}-50.5            \\
                                                                               &                                                           & NDCG@5           &   \cellcolor[HTML]{FFCCC9}-31.0                            &+36.3                          &\cellcolor[HTML]{FFCCC9}-41.2                               &\cellcolor[HTML]{FFCCC9}-35.7            &\cellcolor[HTML]{FFCCC9}-13.0            &\cellcolor[HTML]{FFCCC9}-9.00                               &\cellcolor[HTML]{FFCCC9}-66.3            &+1.20            &\cellcolor[HTML]{FFCCC9}-14.3                               &\cellcolor[HTML]{FFCCC9}-53.6            \\
                                                                               & \multirow{-3}{*}{ReLLA}                                   & MRR@10              & \cellcolor[HTML]{FFCCC9}-28.9                              &+49.6                          &\cellcolor[HTML]{FFCCC9}-27.7                               &\cellcolor[HTML]{FFCCC9}-24.1            &\cellcolor[HTML]{FFCCC9}-2.80            &\cellcolor[HTML]{FFCCC9}-9.80                               &\cellcolor[HTML]{FFCCC9}-64.4            &\cellcolor[HTML]{FFCCC9}-3.50            &\cellcolor[HTML]{FFCCC9}-15.7                               &\cellcolor[HTML]{FFCCC9}-41.4            \\ \cline{2-13} 
                                                                               &                                                           & HR@5             &   \cellcolor[HTML]{FFCCC9}-38.8                            &+48.3                          &\cellcolor[HTML]{FFCCC9}-41.2                               &\cellcolor[HTML]{FFCCC9}-33.8            &\cellcolor[HTML]{FFCCC9}-10.1            &\cellcolor[HTML]{FFCCC9}-4.40                               &\cellcolor[HTML]{FFCCC9}-51.3            &+6.30            &\cellcolor[HTML]{FFCCC9}-6.20                               &\cellcolor[HTML]{FFCCC9}-53.5            \\
                                                                               &                                                           & NDCG@5           & \cellcolor[HTML]{FFCCC9}-46.9                              &+57.2                          &\cellcolor[HTML]{FFCCC9}-43.6                               &\cellcolor[HTML]{FFCCC9}-42.4            &\cellcolor[HTML]{FFCCC9}-16.1            &\cellcolor[HTML]{FFCCC9}-6.90                               &\cellcolor[HTML]{FFCCC9}-62.9            &\cellcolor[HTML]{FFCCC9}-4.90            &\cellcolor[HTML]{FFCCC9}-7.40                               &\cellcolor[HTML]{FFCCC9}-54.1            \\
                                                                               & \multirow{-3}{*}{CALRec}                                  & MRR@10              &\cellcolor[HTML]{FFCCC9}-41.5                               &+61.0                          &\cellcolor[HTML]{FFCCC9}-29.6                               &\cellcolor[HTML]{FFCCC9}-30.6            &\cellcolor[HTML]{FFCCC9}-5.10            &\cellcolor[HTML]{FFCCC9}-7.40                               &\cellcolor[HTML]{FFCCC9}-60.2            &\cellcolor[HTML]{FFCCC9}-7.20            &\cellcolor[HTML]{FFCCC9}-6.10                               &\cellcolor[HTML]{FFCCC9}-42.0            \\ \cline{2-13} 
                                                                               &                                                           & HR@5             &  +14.8                             &+41.1                          &+10.2                               &+6.20            &+29.1            &+0.30                               &+1.10            &+36.0            &+11.5                               &+14.7            \\
                                                                               &                                                           & NDCG@5           & +4.4                              &+49.9                          &+9.70                               &+1.00            &+29.1            &+1.10                               &\cellcolor[HTML]{FFCCC9}-5.40            &+46.3            &+23.3                               &+23.3            \\
\multirow{-12}{*}{\textbf{\begin{tabular}[c]{@{}c@{}}LLM\\ -Rec\end{tabular}}} & \multirow{-3}{*}{\textbf{Ours}} & MRR@10 & +2.9 & +55.4 & +20.5 & +9.70 & +37.5 & +1.50 & \cellcolor[HTML]{FFCCC9}-6.30 & +47.3 & +27.6 & +33.6 \\

\bottomrule
\end{tabular}
  \begin{tablenotes}
    \item[*] *\textbf{Amazon}: \textbf{B}($Books$), \textbf{G}($Games$), \textbf{J}($Jewelry$), \textbf{O}($Outdoors$), \textbf{T}($Toys$) \\ * \textbf{Telco}:  \textbf{W}($Web$-$View$), \textbf{C}($Call$-$Log$), \textbf{P}($PoI$), \textbf{B}($Benefits$), \textbf{S}($Shopping$)
    \end{tablenotes}
\end{table}

\subsection{Cold \& Warm Scenario (RQ3)} \label{subsection:diss_cold_warm}
\subsubsection{Cold/Warm Item Scenario}

\begin{figure}[htbp]
\centering
\includegraphics[width=0.94\linewidth]{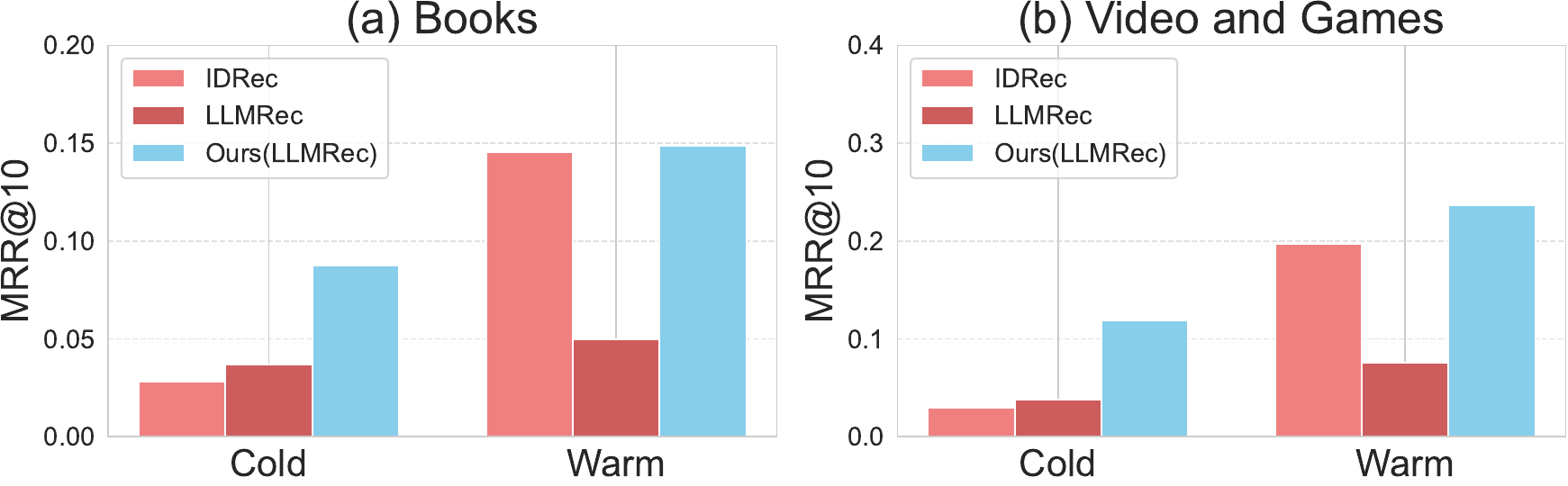}
\caption{Performance Comparison with LLMRec and IDRec Models under Cold/Warm-Item Scenario}
\label{fig:cold_warm_reslut}
\end{figure}

Following \citet{kim2024large}, we defined cold items as those in the bottom 10\% (\textbf{Amazon}) or 30\% (\textbf{Telco}) of interaction frequency, and warm items as all remaining items.
We then compared performance in cases where the ground-truth item is either cold or warm. 
As shown in Fig. \ref{fig:cold_warm_reslut}, LLMRec models showed lower performance on warm-items but higher performance on cold-items compared to IDRec models such as SASRec~\cite{kang2018self}. 
In general, in warm-item scenarios with many user-item interactions, LLMRec models tend to perform worse than simpler traditional collaborative filtering models (i.e., IDRec)~\cite{kim2024large} due to its heavy reliance on textual information.
\textbf{In contrast, our model successfully captured item-level collaborative signals and achieved superior performance not only on cold-items but also on warm-items compared to IDRec models}.
This result indicates a significant extension of the applicability of the model, underlining its practical utility in real-world applications.
The results for all domains are provided in Appendix~\ref{section: extended_cold_and_warm}.

\subsubsection{Cold User Scenario}

\begin{figure}[htbp]
\centering
\includegraphics[width=0.94\linewidth]{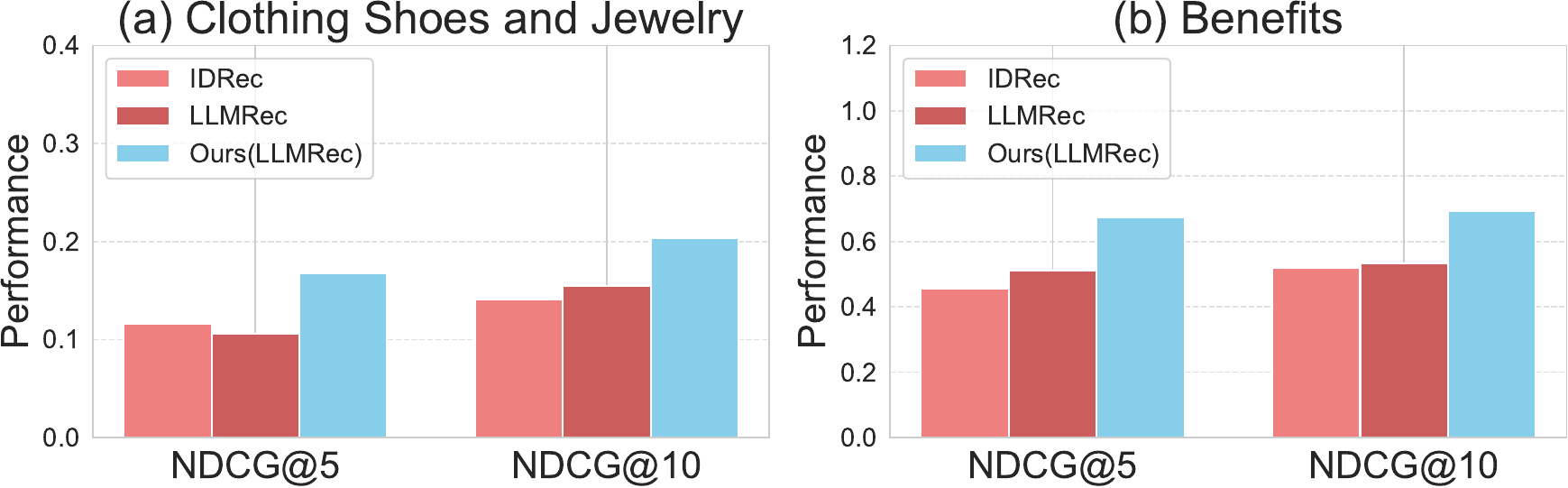}
\caption{Performance Comparison under Cold-User Scenario}
\label{fig:cold_user}
\end{figure}

We additionally constructed a cold-user evaluation setting, defining cold users as those in the bottom 1\% of the interaction length distribution across all users.
Following \citet{kim2024large}, the model was trained exclusively on the remaining users and evaluated on this cold-user subset.
Across domains, DuELRec consistently outperformed IDRec models, with particularly notable gains in domains where sparse collaborative knowledge limits performance for cold users (Fig.~\ref{fig:cold_user}).
This consistent advantage stems from our model’s ability to leverage LLMs to capture semantic user characteristics, enabling higher performance even for cold users with limited collaborative knowledge.
Similar trends were observed across most domains; see Appendix~\ref{section: extended_cold_and_warm} for details.

\subsection{Discussion of Model Variants (RQ4)} \label{subsection: discussion_model_variants}

\begin{table}[htbp] \footnotesize
\centering
\caption{Comparison of MRR between different component combinations, where positive gains (\%) of the variants over our model are highlighted in \setlength{\fboxsep}{0pt}\colorbox[HTML]{DAE8FC}{blue} and negative ones in \setlength{\fboxsep}{0pt}\colorbox[HTML]{FFCCC9}{red}.}
    \renewcommand{\arraystretch}{1.2}
\label{tab:result_4}
\setlength{\tabcolsep}{0.8pt}
\begin{tabular}{l|ccccc|ccccc}
\toprule
\multicolumn{1}{c|}{\cellcolor[HTML]{EFEFEF}\textbf{Dataset}}  & \multicolumn{5}{c|}{\cellcolor[HTML]{EFEFEF}\textbf{Amazon}}                                                                                                 & \multicolumn{5}{c}{\cellcolor[HTML]{EFEFEF}\textbf{Telco}}                                                                                                 \\ \hline
\multicolumn{1}{c|}{\textbf{Variants}} & B                        & G                        & J                        & O                        & T                        & W                        & C                        & P                        & B                       & S                        \\ \hline
(A) \textit{w/o} \texttt{DGDE} & \cellcolor[HTML]{DAE8FC}+1.87\% & \cellcolor[HTML]{FFCCC9}-0.23\% & \cellcolor[HTML]{FFCCC9}-22.2\% & \cellcolor[HTML]{FFCCC9}-11.3\% & \cellcolor[HTML]{FFCCC9}-8.00\% & \cellcolor[HTML]{FFCCC9}-4.73\% & \cellcolor[HTML]{FFCCC9}-49.0\% & \cellcolor[HTML]{FFCCC9}-21.4\% & \cellcolor[HTML]{FFCCC9}-21.0\% & \cellcolor[HTML]{FFCCC9}-36.4\% \\
(B) \textit{w/o} \texttt{SE} & \cellcolor[HTML]{FFCCC9}-17.9\% & \cellcolor[HTML]{FFCCC9}-10.4\% & \cellcolor[HTML]{FFCCC9}-6.51\% & \cellcolor[HTML]{FFCCC9}-2.59\% & \cellcolor[HTML]{FFCCC9}-5.40\% & \cellcolor[HTML]{FFCCC9}-2.51\% & \cellcolor[HTML]{FFCCC9}-2.84\% & \cellcolor[HTML]{FFCCC9}-1.32\% & \cellcolor[HTML]{FFCCC9}-2.08\% & \cellcolor[HTML]{FFCCC9}-6.05\% \\
(C) \textit{w/o} \texttt{CE} & \cellcolor[HTML]{FFCCC9}-5.15\% & \cellcolor[HTML]{FFCCC9}-1.85\% & \cellcolor[HTML]{DAE8FC}0.94\% & \cellcolor[HTML]{FFCCC9}-1.94\% & \cellcolor[HTML]{FFCCC9}-0.33\% & \cellcolor[HTML]{FFCCC9}-0.28\% & \cellcolor[HTML]{FFCCC9}-2.19\% & \cellcolor[HTML]{DAE8FC}+1.03\% & \cellcolor[HTML]{FFCCC9}-1.02\% & \cellcolor[HTML]{FFCCC9}-3.29\% \\
(D) \textit{w/o} \texttt{SSCN} & \cellcolor[HTML]{FFCCC9}-0.48\% & \cellcolor[HTML]{DAE8FC}+9.94\% & \cellcolor[HTML]{FFCCC9}-35.4\% & \cellcolor[HTML]{FFCCC9}-13.3\% & \cellcolor[HTML]{FFCCC9}-10.1\% & \cellcolor[HTML]{FFCCC9}-6.36\% & \cellcolor[HTML]{FFCCC9}-53.3\% & \cellcolor[HTML]{FFCCC9}-19.2\% & \cellcolor[HTML]{FFCCC9}-12.9\% & \cellcolor[HTML]{FFCCC9}-41.5\% \\
(E) \textit{w/} \texttt{IDEmb}  & \cellcolor[HTML]{FFCCC9}-20.4\% & \cellcolor[HTML]{DAE8FC}+3.01\% & \cellcolor[HTML]{FFCCC9}-31.2\% & \cellcolor[HTML]{DAE8FC}+0.56\% & \cellcolor[HTML]{FFCCC9}-29.3\% & \cellcolor[HTML]{FFCCC9}-11.9\% & \cellcolor[HTML]{FFCCC9}-59.9\% & \cellcolor[HTML]{FFCCC9}-39.1\% & \cellcolor[HTML]{DAE8FC}+17.6\% & \cellcolor[HTML]{FFCCC9}-233.6\% \\ 
\bottomrule
\end{tabular}
\end{table}

\subsubsection{Component Analysis} To validate the effectiveness of each component, we propose several variants as follows.

\noindent \textbf{(A) \textit{w/o} \texttt{Dual-Expert}}: The \textit{Domain-Gated Dual Sequential Experts} (\texttt{DGDE}) module (Section \ref{subsection:dgde}) was replaced with the basic transformer~\cite{vaswani2017attention} in this model variant. 
The absence of \texttt{DGDE} indicates that the LLM fails to capture item-level collaborative signals.

\noindent \textbf{(B) \textit{w/o} \texttt{Single-Expert}}: Only the cross-domain expert was retained in the \texttt{DGDE}, forcing the model to rely solely on cross-domain information without capturing domain-specific sequential patterns.

\noindent \textbf{(C) \textit{w/o} \texttt{Cross-Expert}}: With the cross-domain expert removed in the \texttt{DGDE}, the model captures domain-specific patterns but cannot leverage cross-domain information.


\noindent \textbf{(D) \textit{w/o} \texttt{SSCN}}: The \textit{Stochastic Single- and Cross-Domain Negative Sampler} (\texttt{SSCN}) was replaced with the negative sampler based on uniform distribution.

\noindent \textbf{(E) \textit{w/} \texttt{IDEmb}}: We tested a DuELRec variant without the LLM backbone, replacing the text input with ID input.

\noindent \textbf{The primary components of our model each play a critical role in enhancing its overall recommendation performance.} 
As shown in Table \ref{tab:result_4}, a comparative analysis of our model and its five variants showed that the removal or substitution of any of the key components led to a significant degradation in performance.
In particular, the difference in performance between our model and the (A) $w/o$ \texttt{DGDE} variant underscores the importance of modeling item-level collaborative signals rather than relying solely on subtoken-level sequential patterns. 
The difference in performance between our model and the (B) $w/o$ \texttt{Single-Expert} variant underscores the importance of capturing domain-specific sequential patterns in addition to shared cross-domain signals.
Similarly, the performance gap between our model and the (C) $w/o$ \texttt{Cross-Expert} variant highlights the necessity of leveraging transferable patterns across domains, rather than relying solely on isolated single-domain information. 
Furthermore, the performance drop in the (D) $w/o$ \texttt{SSCN} variant compared to our model in most domains highlights the effectiveness of the \texttt{SSCN} module in capturing transferable item correlations both within and across domains, and thus mitigating negative transfer between domains, with a detailed sensitivity analysis on $p$ provided in Section~\ref{subsection:hyper_ablation}.
Meanwhile, despite marginal gains in a few domains, the (E) $w/$ \texttt{IDEmb} variant underperformed overall, highlighting the benefits of our ID-free design in performance, cold-start handling, and deployment simplicity.

\subsubsection{Analysis on Instruction-Tuning}

\begin{figure}[htbp]
\centering
\includegraphics[width=0.94\linewidth]{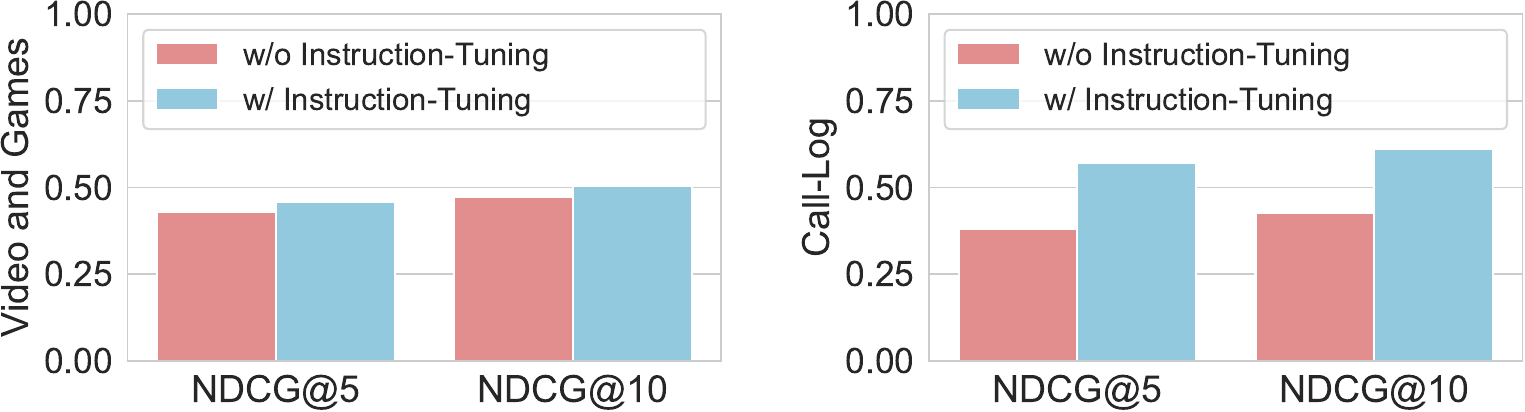}
\caption{Domain-Wise Performance Comparison: Without vs. With Instruction-Tuning (\textit{Games} and \textit{Call}-\textit{Log} Domains)}
\label{fig:inst_tuning}
\end{figure}

In Section \ref{subsection:sft}, we proposed an auxiliary task (i.e., $\mathcal{L}_{inst}$) that allows the LLM backbone to better adapt to textual user behavior sequences with recommendation instructions.
We conducted experiments to compare performance with and without this auxiliary task (Fig.~\ref{fig:inst_tuning}).
The auxiliary task with PEFT significantly improved performance, as shown in Fig.~\ref{fig:inst_tuning}.
This shows that allowing the LLM to learn from user behavior logs effectively improves recommendation performance.
Detailed results for all domains are provided in Appendix~\ref{section: extended_inst_tuning}.

\subsection{Sensitivity Analysis (RQ5)}  \label{subsection:hyper_ablation}

\begin{figure}[htbp]
\centering
\includegraphics[width=0.99\linewidth]{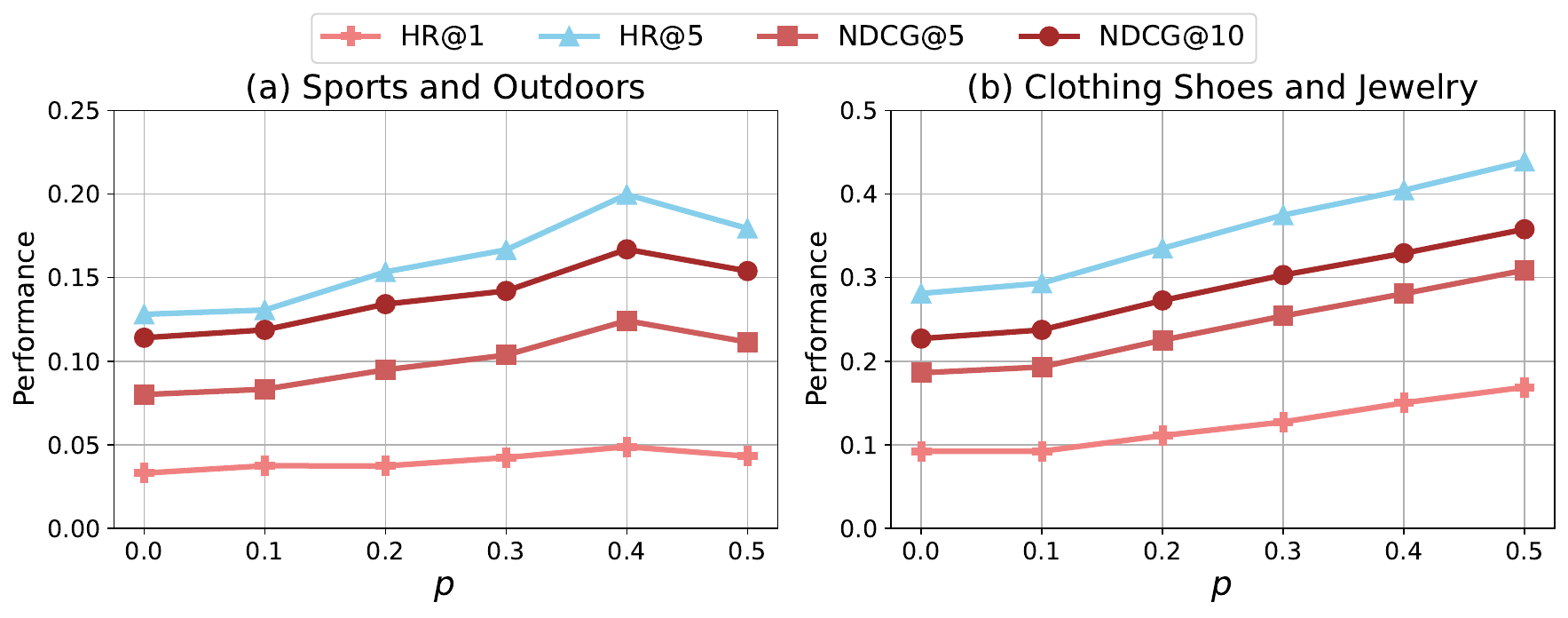}
\caption{Sensitivity analysis of $p$ for negative sampling.}
\label{fig:sensitivity}
\end{figure}

We conducted a sensitivity analysis on the hyperparameter \(p\), which controls the proportion of cross-domain negative samples during contrastive learning, as shown in Fig.~\ref{fig:sensitivity}.
We varied \(p\) from 0 to 0.5 in increments of 0.1 while keeping all other settings fixed. When \(p = 0\), negatives were sampled exclusively from the same domain---a setting commonly adopted in prior studies---resulting in noticeably lower performance. In contrast, introducing cross-domain negatives (\(p > 0\)) consistently improved both accuracy and robustness. 
This performance gain becomes more pronounced as \(p\) increases, indicating that our model effectively suppresses negative transfer and instead leverages cross-domain exposure to amplify positive transfer, resulting in robust performance improvements.
Due to space limitations, we report results for only two representative domains, which mirror the trends observed across all domains; see Appendix~\ref{section: extended_sensitivity} for the complete results.

\subsection{Time Complexity Analysis (RQ6)} 
\subsubsection{Theoretical Analysis}
Let $L$ be the subtoken sequence length, $n$ the item-level sequence length, $d$ the hidden dimension, $H_{\text{LLM}}$ and $H_{\text{EXP}}$ the number of transformer layers in the LLM backbone and each expert, respectively, $K$ the number of negative samples in the training stage, and $N$ the number of item candidates in the inference.  
During training, DuELRec processes length-$L$ sequences with a frozen LLM backbone ($\mathcal{O}(H_{\text{LLM}}L^2d)$) and two shallow experts ($\mathcal{O}(2H_{\text{EXP}}L^2d)$), combines them via gating ($\mathcal{O}(nd)$), and applies a contrastive loss over $K$ negatives ($\mathcal{O}(Kd)$), resulting in a total complexity of $\mathcal{O}\big(L^2 d (H_{\text{LLM}} + 2 H_{\text{EXP}}) + Kd\big)$.  
During inference, the contrastive step is replaced by similarity computation over $N$ candidates ($\mathcal{O}(Nd)$), giving a complexity of $\mathcal{O}\big(L^2 d (H_{\text{LLM}} + 2 H_{\text{EXP}}) + Nd\big)$.

\subsubsection{Empirical Analysis} 
Compared to ID-based models (e.g., SyNCRec), DuELRec is on average \textbf{1.67× slower} in training and \textbf{2.29× slower} in inference; however, it requires no retraining for new items and achieves superior cross-domain performance (Tables~\ref{tab:result_1} and \ref{tab:result_2}).  
Against other LLM-based models (e.g., RecFormer), DuELRec is \textbf{1.16× faster} in training and \textbf{3.19× faster} in inference.

DuELRec’s efficiency stems from three design choices:
(i) it uses only two experts regardless of the number of domains, ensuring scalability---unlike SyNCRec, which allocates one expert per domain, or BiTGCF and CMF, which require modeling all ${|\mathcal{D}| \choose 2}$ domain pairs;  
(ii) it reduces the sequence length from $L$ to $n$ after the LLM backbone, making contrastive learning more efficient than LLMRec models that operate directly on $L$; and  
(iii) it adopts a single-stage training pipeline, in contrast to the two-stage (pretraining–finetuning) approach of RecFormer and SAID, further improving training efficiency.

\subsection{Online A/B Test (RQ7)}  \label{subsection:ab_test}

Our recommendation model was deployed in the personal assistant app of a major telecommunications company and evaluated through a large-scale online A/B test conducted from January to March 2025.
To meet business constraints such as latency tolerance, we implemented a customized version of the model trained on sequential interaction logs from approximately 5 million users across 14 domains, including the app’s target domain.
The task was to recommend the Top-1 next item within the target domain based on each user’s full cross-domain interaction history.
Traffic was randomly split throughout the evaluation period in a fixed 45:35:20 ratio among three model buckets: a popularity-based baseline, a SASRec variant~\cite{park2023cracking}, and our proposed model.
Click-through rate (CTR), defined as the ratio of clicks to total impressions, was used as the primary evaluation metric.
As shown in Table~\ref{tab:online_ab}, our model achieved a CTR of 1.33\%, significantly outperforming the SASRec variant (0.90\%) and the popularity baseline (0.63\%).
A two-tailed two-proportion z-test confirmed both improvements as highly significant ($p < 0.01$).
These results demonstrate the practical effectiveness of our model in real-world deployment, with latency performance suitable for production environments.

\begin{table}[htbp] \small
\centering
\caption{Online A/B Test Results: Performance Comparison}
    \renewcommand{\arraystretch}{1.1}
\label{tab:online_ab}
\setlength{\tabcolsep}{1.8pt}
\begin{tabular}{c|c|c|c|c}
\toprule
\rowcolor[HTML]{EFEFEF} 
\textbf{Model}       & \textbf{\# Impression} & \textbf{\# Clicks} & \textbf{CTR} & \textbf{Inference Latency} \\ \hline
\textbf{Popularity}       & 16.31M              & 103K          &  0.63\% &    ~50ms/call               \\
\textbf{SASRec Variant}      &   13.95M            &  126K         &  0.90\% &   ~105ms/call              \\
\textbf{Ours}             &  5.66M             &   75K        &  1.33\% &    ~242ms/call               \\ 
\bottomrule
\end{tabular}
\end{table}

\section{Conclusion} \label{section: future_work}
We propose DuELRec, a domain-gated dual-expert framework with LLMs, to address negative transfer in cross-domain sequential recommendation. By fusing item-level collaborative signals through expert gating and dual-sampling contrastive learning, DuELRec captures both single- and cross-domain patterns. Extensive experiments and real-world deployment demonstrate its effectiveness.

\begin{acks}
We gratefully acknowledge SK Telecom for providing the GPU cluster used in this research.
This work was supported by Institute for Information \& communications Technology Planning \& Evaluation(IITP) grant funded by the Korea government(MSIT) (RS-2019-II190075, Artificial Intelligence Graduate School Program(KAIST)). This work was also supported by the National Research Foundation of Korea(NRF) grant funded by the Korea government(MSIT) (No. RS-2025-00555621).
\end{acks}

\clearpage
\section*{GenAI Usage Disclosure}
We used large language models (e.g., ChatGPT) solely for language polishing to improve the clarity and readability of this paper. No GenAI tools were used for ideation, experimental design, data analysis, or result generation.
\balance
\bibliographystyle{ACM-Reference-Format}
\bibliography{custom}

\clearpage
\appendix

\noindent\textbf{A Dual-Expert Strategy Integrating LLMs to Mitigate Negative Transfer in Cross-Domain Sequential Recommendation}

\noindent\textbf{Appendix}

\section{Baselines} \label{section:baselines}
In this section, we review key previous studies that have addressed the CDSR task (baselines compared in Tables~\ref{tab:result_1},~\ref{tab:result_2}, ~\ref{tab:pair_result}, and ~\ref{tab:pair_result_cdsr}).
\subsection{IDRec for CDSR}
CGRec~\cite{park2023cracking} and SyNCRec~\cite{park2024pacer} represent pioneering studies in Cross-Domain Sequential Recommendation (CDSR) as all-in-one models capable of learning from three or more domains simultaneously. In contrast, methods such as MAN~\cite{lin2023mixed}, C2DSR~\cite{cao2022contrastive}, $\pi$-Net and MIFN~\cite{ma2022mixed} are specifically designed to address CDSR challenges within two-domain pairs, limiting their applicability to multi-domain scenarios.
For this reason, we present a detailed performance result of these models, which are limited to handling two-domain pairs, separately in Table  \ref{tab:pair_result} and \ref{tab:pair_result_cdsr} , as opposed to the main performance results discussed in Table \ref{tab:result_1} and \ref{tab:result_2}.

\begin{itemize}
  \item \textbf{CGRec}~\cite{park2023cracking} addresses the challenge of negative transfer in CDSR by proposing a framework that adaptively reduces the influence of dissimilar domains using cooperative game theory to assess domain contributions. Additionally, a hierarchical contrastive learning approach leveraging coarse- and fine-level category information enhances model performance, achieving state-of-the-art results on real-world datasets.
  
  \item \textbf{SyNCRec}~\cite{park2024pacer} also tackles negative transfer in CDSR by introducing an adaptive weighting mechanism based on the degree of negative transfer and leveraging mutual information between CDSR and SDSR tasks.

  \item \textbf{C$^{2}$DSR}~\cite{cao2022contrastive} simultaneously captured single- and cross-domain user preferences by utilizing a graph neural network and a sequential attentive encoder to model intra- and inter-sequence item relationships. 
  It further incorporated a contrastive cross-domain infomax objective to strengthen the connection between single- and cross-domain user representations.

  \item \textbf{MoSE}~\cite{qin2020multitask} addresses the challenge of modeling sequential user behavior in multi-task learning settings, incorporating Long Short-Term Memory (LSTM) into the Multi-gate Mixture-of-Experts architecture. 
  It effectively captures temporal dependencies and handles heterogeneous data sources like web search and browsing logs with differing properties. 
  
  \item \textbf{MAN}~\cite{lin2023mixed} addresses data sparsity in sequential recommendation by leveraging cross-domain information without relying on overlapped users. It employs local and global attention modules to capture domain-specific and domain-general patterns, utilizing a mixed attention layer and prediction layer to effectively fuse and evolve user interests, demonstrating superior performance on real-world datasets.

  \item 
  \textbf{$\pi$-Net}~\cite{ma2019pi} addresses the Shared-account Cross-domain Sequential Recommendation (SCSR) task by introducing a Parallel Information-sharing Network that simultaneously generates recommendations for two domains. 
  It employs a Cross-Domain Transfer Unit (CTU) to adaptively transfer information across domains, achieving improved recommendations by integrating cross-domain user behavior. 

  \item \textbf{MIFN}~\cite{ma2022mixed} enhances the performance of CDSR task by modeling both the flow of behavioral information and knowledge across domains using behavior and knowledge transfer units. 
  This mixed information flow network selectively utilizes cross-domain information based on users’ current preferences, achieving improved performance on e-commerce datasets.
  
\end{itemize}

\subsection{LLMRec}
LLMRec is a rapidly emerging field of research that integrates the semantic features of user behavior sequences, represented in textual form, with the open-world knowledge of Large Language Models (LLMs). As a result, LLMRec demonstrates competitive performance even in cross-domain settings.

\begin{itemize}
  \item \textbf{UniSRec}~\cite{hou2022towards} utilizes pre-trained language models to encode item texts and introduces a parametric whitening and mixture-of-experts (MoE) adaptors to create isotropic, transferable item representations for sequential recommendation.

\item \textbf{E4SRec}~\cite{li2023e4srec} focuses on a lightweight integration of LLMs with traditional ID-based recommendation systems, using minimal pluggable components to enhance efficiency and scalability. 

\item \textbf{RecFormer}~\cite{li2023text} introduces a framework for ID-free sequential recommendation by learning language representations through a bi-directional Transformer model that encodes item key-value attributes as text.
  This approach improves transferability across domains and cold-start settings, demonstrating significant performance gains in both supervised and zero-shot recommendations.

\item \textbf{SAID}~\cite{hu2024enhancing} proposes a two-step approach to sequential recommendation. In the first step, a projector module maps item IDs into embeddings that are semantically compatible with the representation space of large language models (LLMs). In the second step, these aligned embeddings are fed into efficient sequential models—such as GRU or Transformer—that are trained to perform next-item prediction.

\item \textbf{CALRec}~\cite{li2024calrec} introduces a two-phase fine-tuning strategy for generative LLMs, combining joint training across multiple categories with specialized tuning for individual categories. The model is trained using a hybrid loss that merges next-item generation with contrastive alignment objectives, and it leverages a quasi-round-robin BM25-based retrieval method to enhance recommendation efficiency.
We performed inference using cosine similarity–based matching in our practical implementation.

\item \textbf{ReLLa}~\cite{lin2024rella} introduces a framework that enhances Large Language Models (LLMs) for recommendation tasks by addressing the challenge of long user behavior sequences using two key components: semantic user behavior retrieval (SUBR) and retrieval-enhanced instruction tuning (ReiT). SUBR improves data quality by selecting semantically relevant behaviors, while ReiT augments training samples with retrieval-enhanced counterparts to increase robustness and generalization.This design enables ReLLa to outperform traditional models even in few-shot settings, leveraging LLMs’ reasoning abilities to comprehend lifelong sequential behavior, making it effective for both zero-shot and few-shot recommendation scenarios.

\item \textbf{Lite-LLMRec}~\cite{wang2024rethinking} proposes a hierarchical LLM structure with two components: an Item LLM to encode item context into compact embeddings and a Recommendation LLM to process these embeddings efficiently. The model replaces beam search decoding with a lightweight item projection head, enabling faster inference while preserving performance. In our implementation, the Item LLM employs a fully-connected layer, and all sub-token embeddings of each item are used as input to the Recommendation LLM.

\item \textbf{LLMEmb}~\cite{liu2024large} incorporates collaborative signals into LLM-generated embeddings by modifying the contrastive loss, improving the model's suitability for recommendation tasks.
 In our implementation, due to computational constraints, we omitted the PCA transformation and the CF-Rec alignment step.

\item \textbf{HLLM}~\cite{chen2024hllm} introduces a two-tier architecture leveraging Large Language Models (LLMs) for sequential recommendation by combining an Item LLM to extract rich content features and a User LLM to model user interests. 
In our implementation, a fully connected layer is applied within the Item LLM, and the complete set of sub-token embeddings for each item is passed to the User LLM as input.
\end{itemize}

\section{Implementation Details} \label{section: hyperparameter_setting}

 CGRec \cite{park2023cracking}\footnote{{\color[HTML]{0037D7}\url{https://github.com/cpark88/CGRec}}}, SyNCRec \cite{park2024pacer}\footnote{{\color[HTML]{0037D7}\url{https://github.com/cpark88/SyNCRec}}}, UniSRec \cite{hou2022towards}\footnote{{\color[HTML]{0037D7}\url{https://github.com/RUCAIBox/UniSRec}}}, and E4SRec \cite{li2023e4srec}\footnote{{\color[HTML]{0037D7}\url{https://github.com/HestiaSky/E4SRec}}} were executed with their official code.
RecFormer ~\cite{li2023text}, Lite-LLMRec ~\cite{wang2024rethinking}, SAID ~\cite{hu2024enhancing}, LLMEmb~\cite{liu2024large}, ReLLa ~\cite{lin2024rella}, A-LLMRec~\cite{kim2024large}, HLLM~\cite{chen2024hllm} and CALRec ~\cite{li2024calrec} were implemented from scratch.
All other CDSR and SR models were implemented using the RecBole framework \cite{zhao2021recbole,recbole[2.0]}.
Most ID-based CDSR models, with the exception of CGRec and SyNCRec, were specifically designed to capture \textbf{only pairwise interactions between two domains.} As a result, these baselines were trained using data from two out of five domains. In these models, each domain was paired with the remaining four domains, and the average performance across these pairings is shown in Table \ref{tab:result_1} and \ref{tab:result_2}.

    \begin{figure}[]
    \centering
    \includegraphics[width=0.999 \linewidth]{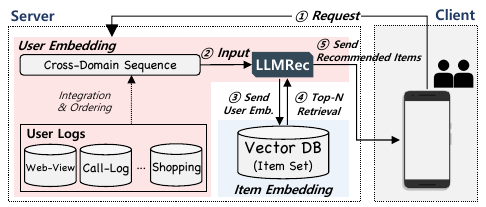}
    \caption{Overview of the inference process}
    \label{fig:inference_process2}
    \end{figure}

\section{The inference and deployment process} \label{section: inf_and_dep}
In this section, we describe the inference process of our proposed model in Fig. \ref{fig:inference_process2}.
$\circled{1}$ The process begins with the client (user side) sending a request to the server, including variables such as \texttt{user\_id}.
On the server side, $\circled{2}$ the user’s cross-domain sequence data is fed into our model to generate a user embedding.
$\circled{3}$ Specifically, the user embedding is derived from the last embedding in the model's output sequence (Section \ref{subsection:train_eval}). 
$\circled{4}$ The user embedding is then compared to a pre-computed set of item embeddings computed with the LLM backbone (Eq. \ref{equation:item_encoder}) and stored in a vector database. 
By performing a similarity calculation, the recommender system identifies the top-$N$ most relevant items for recommendation.
$\circled{5}$ These recommended items are subsequently sent back to the client, providing personalized recommendation results to the user.

\begin{table*}[] \footnotesize
\centering
\caption{Performance comparison of CDR baselines (BiGCF, DTCDR, CMF, DeepAPF) for two paired domains.
}
    \renewcommand{\arraystretch}{1.15}
\label{tab:pair_result}
\setlength{\tabcolsep}{0.79pt}
\begin{tabular}{ccc|cccccc|cccccc|cccccc|cccccc}
\toprule
\multicolumn{3}{c|}{\cellcolor[HTML]{EFEFEF}\textbf{Model}}                                                                                                                                                                                              & \multicolumn{6}{c|}{\cellcolor[HTML]{EFEFEF}\textbf{BiGCF}}        & \multicolumn{6}{c|}{\cellcolor[HTML]{EFEFEF}\textbf{DTCDR}}        & \multicolumn{6}{c|}{\cellcolor[HTML]{EFEFEF}\textbf{CMF}}        & \multicolumn{6}{c}{\cellcolor[HTML]{EFEFEF}\textbf{DeepAPF}}        \\ \hline
\multicolumn{1}{c|}{}                                    & \multicolumn{1}{c|}{}                                                                                   &                                                                                     & \multicolumn{3}{c}{HR}        & \multicolumn{2}{c}{NDCG} & MRR     & \multicolumn{3}{c}{HR}        & \multicolumn{2}{c}{NDCG} & MRR     & \multicolumn{3}{c}{HR}      & \multicolumn{2}{c}{NDCG} & MRR     & \multicolumn{3}{c}{HR}         & \multicolumn{2}{c}{NDCG} & MRR     \\ \cline{4-27} 
\multicolumn{1}{c|}{\multirow{-2}{*}{\textbf{Dataset}}}           & \multicolumn{1}{c|}{\multirow{-2}{*}{\textbf{\begin{tabular}[c]{@{}c@{}}Target\\ Domain\end{tabular}}}} & \multirow{-2}{*}{\textbf{\begin{tabular}[c]{@{}c@{}}Source \\ Domain\end{tabular}}} & @1       & @5       & @10     & @5          & @10        & @10     & @1       & @5       & @10     & @5          & @10        & @10     & @1      & @5      & @10     & @5          & @10        & @10     & @1       & @5       & @10      & @5          & @10        & @10     \\ \hline
\multicolumn{1}{c|}{}                                    & \multicolumn{1}{c|}{}                                                                                   & Games                                                                               & 0.152    & 0.332    &0.424   &0.245       &0.275      &0.229   & 0.145    &0.288    &0.368   & 0.219      &0.245      &0.207   &0.165   &0.328   &0.411   &0.250       &0.277      &0.235   &0.146    &0.290    &0.365    &0.221       &0.245      &0.208   \\
\multicolumn{1}{c|}{}                                    & \multicolumn{1}{c|}{}                                                                                   & Jewelry                                                                            &0.124          &0.266          &0.351         &0.198             &0.225            &0.186         &0.144          &0.288          &0.369         &0.219             &0.245            &0.207         &0.162         &0.323         &0.402         &0.246             &0.272            &0.231         &0.142          &0.282          &0.351          &0.215             &0.237            &0.202         \\
\multicolumn{1}{c|}{}                                    & \multicolumn{1}{c|}{}                                                                                   & Outdoors                                                                              &0.155          &0.322          &0.414         &0.242             &0.271            &0.227         &0.143          &0.288          &0.367         &0.219             &0.244            &0.206         &0.161         &0.315         &0.391         &0.242             &0.266            &0.227         &0.134          &0.267          &0.336          &0.203             &0.226            &0.191         \\
\multicolumn{1}{c|}{}                                    & \multicolumn{1}{c|}{\multirow{-4}{*}{\textbf{Books}}}                                                   & Toys                                                                                &0.160          &0.329          &0.419         &0.248             &0.277            &0.233         &0.144          &0.293          &0.371         &0.222             &0.247            &0.209         &0.160         &0.308         &0.386         &0.237             &0.262            &0.224         &0.140          &0.276          &0.346          &0.211             &0.234            &0.199         \\ \cline{2-27} 
\multicolumn{1}{c|}{}                                    & \multicolumn{1}{c|}{}                                                                                   & Books                                                                               &0.173          &0.384          &0.512         &0.281             &0.323            &0.264         &0.151          &0.397          &0.537         &0.277             &0.322            &0.256        &0.142         &0.316         &0.429         &0.231             &0.267            &0.218         &0.144          &0.317          &0.427          &0.233             &0.268            &0.220         \\
\multicolumn{1}{c|}{}                                    & \multicolumn{1}{c|}{}                                                                                   & Jewelry                                                                               &0.166          &0.367          &0.492         &0.268             &0.309            &0.253         &0.149          &0.394          &0.539         &0.275             &0.321            &0.255         &0.139         &0.314         &0.430         &0.228             &0.266            &0.216         &0.151          &0.328          &0.438          &0.242             &0.277            &0.228         \\
\multicolumn{1}{c|}{}                                    & \multicolumn{1}{c|}{}                                                                                   & Outdoors                                                                              &0.168          &0.373          &0.498         &0.273             &0.313            &0.257         &0.150          &0.394          &0.536         &0.275             &0.321            &0.255         &0.141         &0.319         &0.436         &0.231             &0.269            &0.218         &0.147          &0.321          &0.432          &0.236             &0.272            &0.223         \\
\multicolumn{1}{c|}{}                                    & \multicolumn{1}{c|}{\multirow{-4}{*}{\textbf{Games}}}                                                   & Toys                                                                                &0.176         &0.388          &0.511         &0.285             &0.325            &0.267         &0.149          &0.393          &0.538         &0.274             &0.321            &0.254         &0.147         &0.329         &0.443         &0.241             &0.277            &0.227         &0.153          &0.333          &0.443          &0.245             &0.281            &0.231         \\ \cline{2-27} 
\multicolumn{1}{c|}{}                                    & \multicolumn{1}{c|}{}                                                                                   & Books                                                                               &0.097          &0.228          &0.316         &0.164             &0.192            &0.155         &0.098          &0.262          &0.363         &0.182             &0.214            &0.169         &0.113         &0.260         &0.347         &0.189             &0.217            &0.177         &0.108          &0.252          &0.337          &0.182             &0.210            &0.170         \\
\multicolumn{1}{c|}{}                                    & \multicolumn{1}{c|}{}                                                                                   & Games                                                                               &0.105          &0.256          &0.348         &0.182             &0.212            &0.170         &0.097          &0.262          &0.362         &0.182             &0.214            &0.169         &0.112         &0.270         &0.364         &0.193             &0.224            &0.180         &0.108          &0.259          &0.351          &0.186             &0.215            &0.174         \\
\multicolumn{1}{c|}{}                                    & \multicolumn{1}{c|}{}                                                                                   & Outdoors                                                                              &0.100          &0.240          &0.328         &0.172             &0.200            &0.161         &0.096          &0.262          &0.362         &0.181             &0.214            &0.168         &0.115         &0.263         &0.351         &0.191             &0.219            &0.179         &0.115          &0.263          &0.347         &0.191             &0.218            &0.179         \\
\multicolumn{1}{c|}{}                                    & \multicolumn{1}{c|}{\multirow{-4}{*}{\textbf{Jewelry}}}                                                & Toys                                                                                & 0.100         &0.238          &0.330         &0.171             &0.200            &0.161         &0.097          &0.263          &0.361         &0.182             &0.214            &0.169         &0.111         &0.259         &0.346         &0.186             &0.215            &0.174         &0.106          &0.250          &0.339          &0.179             &0.208            &0.168         \\ \cline{2-27} 
\multicolumn{1}{c|}{}                                    & \multicolumn{1}{c|}{}                                                                                   & Books                                                                               &0.102          &0.236          &0.329         &0.170             &0.200            &0.161         &0.112          &0.287          &0.394         &0.202             &0.236            &0.188         &0.103         &0.245         &0.337         &0.175             &0.205            &0.165         &0.096          &0.236          &0.330          &0.167             &0.198            &0.157         \\
\multicolumn{1}{c|}{}                                    & \multicolumn{1}{c|}{}                                                                                   & Games                                                                               &0.115          &0.273          &0.371         &0.196             &0.228            &0.184         &0.112          &0.287          &0.392         &0.202             &0.236            &0.188         &0.111         &0.271         &0.368         &0.193             &0.224            &0.180         &0.109          &0.269          &0.363          &0.191             &0.222            &0.178         \\
\multicolumn{1}{c|}{}                                    & \multicolumn{1}{c|}{}                                                                                   & Jewelry                                                                            &0.094          &0.225          &0.313         &0.161             &0.189            &0.151         &0.113          &0.285          &0.391         &0.201             &0.235            &0.187         &0.106         &0.245         &0.336         &0.177             &0.206            &0.167         &0.105          &0.245          &0.334          &0.177             &0.206            &0.166         \\
\multicolumn{1}{c|}{}                                    & \multicolumn{1}{c|}{\multirow{-4}{*}{\textbf{Outdoors}}}                                                  & Toys                                                                                &0.104          &0.247          &0.342         &0.177             &0.208            &0.167         &0.112          &0.285          &0.393         &0.201             &0.235            &0.187         &0.105         &0.251         &0.346         &0.179             &0.210            &0.168         &0.101          &0.252          &0.346          &0.178             &0.208            &0.166         \\ \cline{2-27} 
\multicolumn{1}{c|}{}                                    & \multicolumn{1}{c|}{}                                                                                   & Books                                                                               &0.109          &0.255          &0.353         &0.183             &0.215            &0.173         &0.110          &0.270          &0.372         &0.192             &0.225            &0.179         &0.099         &0.244         &0.350         &0.172             &0.206            &0.163         &0.092         &0.232          &0.337          &0.162             &0.196            &0.153         \\
\multicolumn{1}{c|}{}                                    & \multicolumn{1}{c|}{}                                                                                   & Games                                                                               & 0.113         &0.269          &0.373         &0.193             &0.226            &0.182         &0.107          &0.270          &0.376         &0.190             &0.225            &0.178         &0.109         &0.265         &0.368         & 0.189            &0.222            &0.177         &0.103          &0.258          &0.359          &0.182             &0.214            &0.170         \\
\multicolumn{1}{c|}{}                                    & \multicolumn{1}{c|}{}                                                                                   & Jewelry                                                                            & 0.104         &0.245          &0.340         &0.176             &0.207            &0.166         &0.108          &0.270          &0.375         &0.191             &0.225            &0.179         &0.099         &0.248         &0.352         &0.175             &0.208            &0.165       &0.093          &0.236          &0.342         &0.166             &0.200            &0.156         \\
\multicolumn{1}{c|}{\multirow{-20}{*}{\textbf{Amazon}}}  & \multicolumn{1}{c|}{\multirow{-4}{*}{\textbf{Toys}}}                                                    & Outdoors                                                                              &0.105          &0.251          &0.351         &0.179             &0.211            &0.169         &0.108          &0.264          &0.373         &0.187             &0.222            &0.177         &0.100         &0.248         &0.353         &0.175             &0.209            &0.165         &0.090          &0.221          &0.313          &0.156             &0.186            &0.148         \\ \hline\hline
\multicolumn{1}{c|}{}                                    & \multicolumn{1}{c|}{}                                                                                   & Call-Log                                                                            &0.849          &0.964          &0.976         &0.915             &0.919            &0.900         &0.717          &0.946          &0.975         &0.846             &0.855            &0.816         &0.794         &0.964         &0.980         &0.892             &0.897            &0.870         &0.839          &0.956          &0.969          &0.908             &0.912            &0.893         \\
\multicolumn{1}{c|}{}                                    & \multicolumn{1}{c|}{}                                                                                   & PoI                                                                                 & 0.845         &0.964          &0.976         &0.915             &0.919            &0.899         &0.717          &0.946          &0.976         &0.846             &0.856            &0.816         &0.818         &0.965         &0.980         &0.903             &0.908            &0.884         &0.844          &0.956          &0.970          &0.910             &0.914            &0.896         \\
\multicolumn{1}{c|}{}                                    & \multicolumn{1}{c|}{}                                                                                   & Benefits                                                                            &0.845          &0.965          &0.977         &0.915             &0.919            &0.899         &0.724          &0.947          &0.976         &0.850             &0.860            &0.821         &0.844         &0.965         &0.977         &0.914             &0.918            &0.898         &0.848          &0.958          &0.971          &0.912             &0.917            &0.898         \\
\multicolumn{1}{c|}{}                                    & \multicolumn{1}{c|}{\multirow{-4}{*}{\textbf{Web-View}}}                                                & Shopping                                                                            &0.847          &0.964          &0.976         &0.915             &0.919            &0.900         &0.717          &0.946          &0.975         &0.846             &0.855            &0.816         &0.830         &0.965         &0.978         &0.909             &0.913            &0.891         &0.850          &0.958          &0.971          &0.913             &0.918            &0.900         \\ \cline{2-27} 
\multicolumn{1}{c|}{}                                    & \multicolumn{1}{c|}{}                                                                                   & Web-View                                                                            &0.334          &0.686          &0.815         &0.520             &0.562            &0.482         &0.334          &0.674          &0.802         &0.514             &0.555            &0.478         &0.420         &0.753         &0.850         &0.599             &0.630            &0.561         &0.446          &0.701          &0.797          &0.584             &0.615            &0.558         \\
\multicolumn{1}{c|}{}                                    & \multicolumn{1}{c|}{}                                                                                   & PoI                                                                                 &0.328          &0.677          &0.809         &0.512             &0.555            &0.475         &0.333          &0.679          &0.813         &0.516             &0.559            &0.480         &0.407         &0.743         &0.846         &0.587             &0.620            &0.549         &0.438          &0.715          &0.809          &0.587             &0.618            &0.557         \\
\multicolumn{1}{c|}{}                                    & \multicolumn{1}{c|}{}                                                                                   & Benefits                                                                            &0.321          &0.670          &0.805         &0.505             &0.549            &0.468         &0.333          &0.675          &0.804         &0.514             &0.556            &0.478         &0.407         &0.743         &0.841         &0.587             & 0.619           &0.548         &0.428          &0.715          &0.817          &0.583             &0.616            &0.552         \\
\multicolumn{1}{c|}{}                                    & \multicolumn{1}{c|}{\multirow{-4}{*}{\textbf{Call-Log}}}                                                & Shopping                                                                            &0.322          &0.669          &0.806         &0.505             &0.549            &0.469         &0.325          &0.669          &0.804         &0.506             &0.550            &0.470         &0.409         &0.738         &0.838         &0.586             &0.618            &0.548         &0.465          &0.733          &0.822          &0.609             &0.638            &0.580         \\ \cline{2-27} 
\multicolumn{1}{c|}{}                                    & \multicolumn{1}{c|}{}                                                                                   & Web-View                                                                            & 0.333         &0.649          &0.777         &0.500             &0.541            &0.467         &0.341          &0.651          &0.786         &0.505             &0.548            &0.474         &0.378         &0.632         &0.765         &0.510             &0.553            &0.487         &0.479          &0.705          &0.801          &0.601             &0.632            &0.579         \\
\multicolumn{1}{c|}{}                                    & \multicolumn{1}{c|}{}                                                                                   & Call-Log                                                                            &0.321          &0.635          &0.773         &0.485             &0.530            &0.454         &0.351          &0.669          &0.801         &0.518             &0.561            &0.486         &0.437         &0.733         &0.827         &0.551             &0.627            &0.564         &0.466          &0.701          &0.797          &0.593             &0.624            &0.570         \\
\multicolumn{1}{c|}{}                                    & \multicolumn{1}{c|}{}                                                                                   & Benefits                                                                            &0.326          &0.652          &0.775         &0.491             &0.535            &0.460         &0.364          &0.682          &0.803         &0.533             &0.572            &0.500         &0.443         &0.726         &0.822         &0.596             &0.627            &0.565         &0.471          &0.704          &0.797          &0.596             &0.626            &0.572         \\
\multicolumn{1}{c|}{}                                    & \multicolumn{1}{c|}{\multirow{-4}{*}{\textbf{PoI}}}                                                     & Shopping                                                                            &0.323          &0.633          &0.771         &0.485             &0.530            &0.455         &0.337          &0.645          &0.783         &0.499             &0.544            &0.469         &0.431         &0.720         &0.817         &0.587             &0.618            &0.555         &0.476          &0.702          &0.792          &0.599             &0.628            &0.628         \\ \cline{2-27} 
\multicolumn{1}{c|}{}                                    & \multicolumn{1}{c|}{}                                                                                   & Web-View                                                                            &0.527          &0.856          &0.925         &0.706             &0.728            &0.665         &0.368          &0.839          &0.932         &0.625             &0.656            &0.566         &0.661         &0.859         &0.933         &0.764             &0.788            &0.743         &0.654          &0.851          &0.921          &0.756             &0.779            &0.734         \\
\multicolumn{1}{c|}{}                                    & \multicolumn{1}{c|}{}                                                                                   & Call-Log                                                                            &0.577          &0.853          &0.915         &0.726             &0.747            &0.693         &0.405          &0.849          &0.938         &0.646             &0.675            &0.590         &0.667         &0.864         &0.931         &0.771             &0.793            &0.749         &0.649          &0.827          &0.899          &0.741             &0.765            &0.723         \\
\multicolumn{1}{c|}{}                                    & \multicolumn{1}{c|}{}                                                                                   & PoI                                                                                 &0.580          &0.862          &0.921         &0.734             &0.753            &0.699         &0.369          &0.833          &0.933         &0.623             &0.656            &0.566         &0.651        &0.859         &0.927         &0.759             &0.781            &0.735         &0.660          &0.806          &0.869          &0.734             &0.755            &0.720         \\
\multicolumn{1}{c|}{}                                    & \multicolumn{1}{c|}{\multirow{-4}{*}{\textbf{Benefits}}}                                                & Shopping                                                                            &0.602          &0.845          &0.899         &0.733             &0.751            &0.704         &0.395          &0.830          &0.927         &0.631             &0.663           &0.578         &0.658         &0.850         &0.914         &0.758             &0.779            &0.736         &0.663          &0.828          &0.880          &0.748             &0.765            &0.729         \\ \cline{2-27} 
\multicolumn{1}{c|}{}                                    & \multicolumn{1}{c|}{}                                                                                   & Web-View                                                                            &0.185          &0.394          &0.527         &0.292             &0.335            &0.276         &0.121          &0.335          &0.480         &0.229             &0.276            &0.214         &0.185         &0.347         &0.488         &0.266             &0.312            &0.258         &0.181          &0.339          &0.481          &0.260             &0.306            &0.253         \\
\multicolumn{1}{c|}{}                                    & \multicolumn{1}{c|}{}                                                                                   & Call-Log                                                                            &0.201          &0.383          &0.503         &0.294             &0.332            &0.280         &0.118          &0.330          &0.478         &0.225             &0.273            &0.211         &0.170         &0.322         &0.472         &0.245             &0.293            &0.239         &0.182          &0.319          &0.450          &0.250             &0.292            &0.244         \\
\multicolumn{1}{c|}{}                                    & \multicolumn{1}{c|}{}                                                                                   & PoI                                                                                 &0.204          &0.373          &0.486         &0.290             &0.326            &0.277         &0.116          &0.339          &0.481         &0.229             &0.275            &0.211         &0.176         &0.357         &0.499         &0.266             &0.312            &0.256         &0.182          &0.333          &0.472          &0.257             &0.302            &0.251         \\
\multicolumn{1}{c|}{\multirow{-20}{*}{\textbf{Telecom}}} & \multicolumn{1}{c|}{\multirow{-4}{*}{\textbf{Shopping}}}                                                & Benefits                                                                            &0.202          &0.372          &0.486         &0.289             &0.326            &0.277         &0.117          &0.343          &0.491         &0.231             &0.279            &0.215         &0.170         &0.363         &0.499         &0.267             &0.311            &0.254         &0.187          &0.353          &0.488          &0.270             &0.314            &0.261         \\ 
\bottomrule
\end{tabular}
\end{table*}

\begin{table*}[] \footnotesize
\centering
\caption{Performance comparison of CDSR baselines (C$^{2}$DSR, MAN, $\pi$-Net, MIFN) for two paired domains.
}
    \renewcommand{\arraystretch}{1.15}
\label{tab:pair_result_cdsr}
\setlength{\tabcolsep}{0.79pt}
\begin{tabular}{ccc|cccccc|cccccc|cccccc|cccccc}
\toprule
\multicolumn{3}{c|}{\cellcolor[HTML]{EFEFEF}\textbf{Model}}                                                                                                                                                                                              & \multicolumn{6}{c|}{\cellcolor[HTML]{EFEFEF}\textbf{C$^{2}$DSR}}        & \multicolumn{6}{c|}{\cellcolor[HTML]{EFEFEF}\textbf{MAN}}        & \multicolumn{6}{c|}{\cellcolor[HTML]{EFEFEF}\textbf{$\pi$-Net}}        & \multicolumn{6}{c}{\cellcolor[HTML]{EFEFEF}\textbf{MIFN}}        \\ \hline
\multicolumn{1}{c|}{}                                    & \multicolumn{1}{c|}{}                                                                                   &                                                                                     & \multicolumn{3}{c}{HR}        & \multicolumn{2}{c}{NDCG} & MRR     & \multicolumn{3}{c}{HR}        & \multicolumn{2}{c}{NDCG} & MRR     & \multicolumn{3}{c}{HR}      & \multicolumn{2}{c}{NDCG} & MRR     & \multicolumn{3}{c}{HR}         & \multicolumn{2}{c}{NDCG} & MRR     \\ \cline{4-27} 
\multicolumn{1}{c|}{\multirow{-2}{*}{\textbf{Dataset}}}           & \multicolumn{1}{c|}{\multirow{-2}{*}{\textbf{\begin{tabular}[c]{@{}c@{}}Target\\ Domain\end{tabular}}}} & \multirow{-2}{*}{\textbf{\begin{tabular}[c]{@{}c@{}}Source \\ Domain\end{tabular}}} & @1       & @5       & @10     & @5          & @10        & @10     & @1       & @5       & @10     & @5          & @10        & @10     & @1      & @5      & @10     & @5          & @10        & @10     & @1       & @5       & @10      & @5          & @10        & @10     \\ \hline
\multicolumn{1}{c|}{}                                    & \multicolumn{1}{c|}{}                                                                                   & Games                                                                               &0.115     &0.292    &0.395   &0.212       &0.238      &0.206   &0.112    &0.254    &0.315   &0.233       &0.253      &0.229   &0.093   &0.203   &0.279   &0.149       &0.174      &0.142   &0.092    &0.199    &0.277    &0.147       &0.172      &0.140   \\
\multicolumn{1}{c|}{}                                    & \multicolumn{1}{c|}{}                                                                                   & Jewelry                                                                            & 0.176         &0.344          &0.434         &0.262             &0.291            &0.264         &0.120          &0.287          &0.361         &0.252             &0.306            &0.274         &0.128         &0.266         &0.351         &0.200             &0.227            &0.189         &0.112          &0.258          &0.348          &0.187             &0.216            &0.176         \\
\multicolumn{1}{c|}{}                                    & \multicolumn{1}{c|}{}                                                                                   & Outdoors                                                                              &0.144          &0.303          &0.414         &0.225             &0.261            &0.233         &0.115          &0.354          &0.400         &0.234             &0.259            &0.249         &0.114         &0.235         &0.322         &0.177             &0.205            &0.169         &0.074          &0.178          &0.222          &0.128             &0.142            &0.117         \\
\multicolumn{1}{c|}{}                                    & \multicolumn{1}{c|}{\multirow{-4}{*}{\textbf{Books}}}                                                   & Toys                                                                                &0.180          &0.340          &0.436         &0.263             &0.294            &0.267         &0.103          &0.367          &0.421         &0.259             &0.276            &0.209         &0.110         &0.239         &0.322         &0.177             &0.203            &0.167         &0.108          &0.239          &0.325          &0.176             &0.204            &0.167         \\ \cline{2-27} 
\multicolumn{1}{c|}{}                                    & \multicolumn{1}{c|}{}                                                                                   & Books                                                                               &0.131          &0.359          &0.494         &0.248             &0.292            &0.250         &0.122          &0.358          &0.407         &0.243             &0.258            &0.267         &0.141         &0.379         &0.505         &0.263             &0.303            &0.241         &0.164          &0.404          &0.535          &0.287             &0.329            &0.266         \\
\multicolumn{1}{c|}{}                                    & \multicolumn{1}{c|}{}                                                                                   & Jewelry                                                                               &0.107          &0.321          &0.465         &0.216             &0.262            &0.222         &0.128          &0.365          &0.412         &0.250             &0.265            &0.273         &0.151         &0.407         &0.548         &0.283             &0.329            &0.261         &0.150          &0.403          &0.543          &0.281             &0.326            &0.260         \\
\multicolumn{1}{c|}{}                                    & \multicolumn{1}{c|}{}                                                                                   & Outdoors                                                                              &0.148          &0.390          &0.527         &0.271             &0.315            &0.269         &0.129          &0.364          &0.509         &0.253             &0.367            &0.277         &0.147         &0.393         &0.530         &0.275             &0.319            &0.254         &0.160          &0.408          &0.541          &0.287             &0.331            &0.266         \\
\multicolumn{1}{c|}{}                                    & \multicolumn{1}{c|}{\multirow{-4}{*}{\textbf{Games}}}                                                   & Toys                                                                                &0.137          &0.376          &0.518         &0.260             &0.306            &0.260         & 0.116         &0.364          &0.519         &0.239             &0.357            &0.263         &0.135         &0.381         &0.511         &0.262             &0.304            &0.240         &0.138          &0.377          &0.507          &0.261             &0.303            &0.240         \\ \cline{2-27} 
\multicolumn{1}{c|}{}                                    & \multicolumn{1}{c|}{}                                                                                   & Books                                                                               &0.096          &0.243          &0.363         &0.169             &0.208            &0.180         &0.118          &0.258          &0.300         &0.138             &0.251            &0.220         &0.085         &0.231         &0.334         &0.160             &0.193            &0.150         &0.088          &0.229          &0.331          &0.160             &0.193            &0.151         \\
\multicolumn{1}{c|}{}                                    & \multicolumn{1}{c|}{}                                                                                   & Games                                                                               &0.080          &0.213          &0.326         &0.146             &0.183            &0.159         &0.051          &0.256          &0.297         &0.163             &0.176            &0.192         &0.073         &0.188         &0.276         &0.132             &0.160            &0.125         &0.095          &0.209          &0.294         &0.154             &0.181            &0.146         \\
\multicolumn{1}{c|}{}                                    & \multicolumn{1}{c|}{}                                                                                   & Outdoors                                                                              &0.097          &0.245          &0.355         &0.172             &0.208            &0.183         &0.096          &0.256          &0.311         &0.135             &0.254            &0.183         &0.089         &0.234         &0.335         &0.163             &0.196            &0.153         &0.104          &0.245          &0.339          &0.175             &0.206            &0.165         \\
\multicolumn{1}{c|}{}                                    & \multicolumn{1}{c|}{\multirow{-4}{*}{\textbf{Jewelry}}}                                                & Toys                                                                                &0.088          &0.234          &0.336         &0.194             &0.225            &0.173         &0.052          &0.272          &0.307         &0.116             &0.228            &0.142         &0.090         &0.226         &0.324         &0.160             &0.191            &0.151         &0.098          &0.235          &0.330          &0.168             &0.198            &0.158         \\ \cline{2-27} 
\multicolumn{1}{c|}{}                                    & \multicolumn{1}{c|}{}                                                                                   & Books                                                                               &0.103          &0.256          &0.373         &0.182             &0.219            &0.192         &0.112          &0.248          &0.390         &0.130             &0.244            &0.154         &0.099         &0.257         &0.352         &0.180             &0.211            &0.211         &0.111          &0.259          &0.353          &0.187             &0.217            &0.176         \\
\multicolumn{1}{c|}{}                                    & \multicolumn{1}{c|}{}                                                                                   & Games                                                                               &0.108          &0.268          &0.380         &0.191             &0.227            &0.199         &0.110          &0.240          &0.378         &0.151             &0.237            &0.153         &0.079         &0.206         &0.286         &0.144             &0.169            &0.134         &0.102          &0.223          &0.298          &0.164             &0.188            &0.155         \\
\multicolumn{1}{c|}{}                                    & \multicolumn{1}{c|}{}                                                                                   & Jewelry                                                                            &0.109          &0.245          &0.352         &0.178             &0.212            &0.191        &0.117          &0.258          &0.295         &0.138             &0.250            &0.159         &0.110         &0.280         &0.378         &0.197             &0.229            &0.183         &0.106          &0.272          &0.371          &0.191             &0.223            &0.178         \\
\multicolumn{1}{c|}{}                                    & \multicolumn{1}{c|}{\multirow{-4}{*}{\textbf{Outdoors}}}                                                  & Toys                                                                                &0.100          &0.253          &0.362         &0.178             &0.213            &0.190         &0.114          &0.206          &0.305         &0.134             & 0.250           &0.159         &0.090         &0.239         &0.333         &0.166             &0.197            &0.155         &0.110          & 0.258         &0.347          &0.186             &0.214            &0.174         \\ \cline{2-27} 
\multicolumn{1}{c|}{}                                    & \multicolumn{1}{c|}{}                                                                                   & Books                                                                               &0.089          &0.254          &0.373         &0.172             &0.211            &0.182         &0.138          &0.268          &0.314         &0.177             &0.292            &0.202         &0.077         &0.220         &0.327         &0.150             &0.184            &0.141         &0.102          &0.246          &0.347          &0.175             &0.208            &0.165         \\
\multicolumn{1}{c|}{}                                    & \multicolumn{1}{c|}{}                                                                                   & Games                                                                               &0.109          &0.281          &0.389         &0.197             &0.232            &0.203         &0.102          &0.260          &0.301         &0.139             &0.252            &0.266         &0.080         &0.203         &0.287         &0.143             &0.170            &0.134         &0.086          &0.206          &0.295          &0.147             &0.176            &0.140         \\
\multicolumn{1}{c|}{}                                    & \multicolumn{1}{c|}{}                                                                                   & Jewelry                                                                            &0.108          &0.267          &0.380         &0.189             &0.225            &0.200         &0.149          &0.273          &0.310         &0.198             &0.209            &0.152         &0.101         &0.265         &0.364         &0.185             &0.217            &0.172         &0.084          &0.241          &0.347          &0.164             &0.198            &0.153         \\
\multicolumn{1}{c|}{\multirow{-20}{*}{\textbf{Amazon}}}  & \multicolumn{1}{c|}{\multirow{-4}{*}{\textbf{Toys}}}                                                    & Outdoors                                                                              &0.097          &0.257          &0.372         &0.179             &0.216            &0.190         &0.100          &0.266          &0.310         &0.168             &0.283            &0.193         &0.099         &0.255         &0.346         &0.179             &0.209            &0.167         &0.095          &0.247          &0.342          &0.173             &0.204            &0.161         \\ \hline\hline
\multicolumn{1}{c|}{}                                    & \multicolumn{1}{c|}{}                                                                                   & Call-Log                                                                            &0.786&	0.947&	0.971&	0.877&	0.885&	0.858&	0.781&	0.924&	0.969&	0.835&	0.880&	0.847&	0.636&	0.896&	0.946&	0.780&	0.796&	0.747&	0.641&	0.898&	0.948&	0.783&	0.799&	0.751         \\
\multicolumn{1}{c|}{}                                    & \multicolumn{1}{c|}{}                                                                                   & PoI                                                                                 &0.776&	0.942&	0.968&	0.869&	0.878&	0.850&	0.736&	0.915&	0.975&	0.846&	0.860&	0.844&	0.623&	0.888&	0.943&	0.769&	0.787&	0.737&	0.624&	0.889&	0.942&	0.770&	0.788&	0.737         \\
\multicolumn{1}{c|}{}                                    & \multicolumn{1}{c|}{}                                                                                   & Benefits                                                                            &0.738&	0.923&	0.956&	0.841&	0.852&	0.714&	0.755&	0.925&	0.968&	0.855&	0.863&	0.706&	0.603&	0.879&	0.937&	0.755&	0.774&	0.722&	0.600&	0.884&	0.939&	0.756&	0.774&	0.721        \\
\multicolumn{1}{c|}{}                                    & \multicolumn{1}{c|}{\multirow{-4}{*}{\textbf{Web-View}}}                                                & Shopping                                                                            &0.754&	0.934&	0.961&	0.853&	0.862&	0.832&	0.751&	0.924&	0.958&	0.850&	0.862&	0.846&	0.592&	0.874&	0.930&	0.746&	0.765&	0.711&	0.586&	0.872&	0.933&	0.744&	0.764&	0.709  \\ \cline{2-27} 
\multicolumn{1}{c|}{}                                    & \multicolumn{1}{c|}{}                                                                                   & Web-View                                                                            &0.466&	0.693&	0.796&	0.585&	0.618&	0.573&	0.382&	0.697&	0.782&	0.575&	0.612&	0.512&	0.451&	0.685&	0.785&	0.572&	0.605&	0.548&	0.553&	0.779&	0.863&	0.675&	0.703&	0.652  \\
\multicolumn{1}{c|}{}                                    & \multicolumn{1}{c|}{}                                                                                   & PoI                                                                                 &0.453&	0.700&	0.808&	0.583&	0.618&	0.567&	0.395&	0.710&	0.793&	0.538&	0.566&	0.515&	0.456&	0.679&	0.782&	0.572&	0.606&	0.551&	0.431&	0.676&	0.784&	0.559&	0.594&	0.535  \\
\multicolumn{1}{c|}{}                                    & \multicolumn{1}{c|}{}                                                                                   & Benefits                                                                            &0.355&	0.621&	0.752&	0.494&	0.536&	0.481&	0.375&	0.631&	0.736&	0.492&	0.507&	0.479&	0.423&	0.653&	0.765&	0.543&	0.579&	0.522&	0.425&	0.655&	0.767&	0.545&	0.582&	0.524  \\
\multicolumn{1}{c|}{}                                    & \multicolumn{1}{c|}{\multirow{-4}{*}{\textbf{Call-Log}}}                                                & Shopping                                                                            &0.480&	0.715&	0.805&	0.603&	0.632&	0.588&	0.474&	0.693&	0.797&	0.525&	0.575&	0.534&	0.413&	0.644&	0.757&	0.534&	0.570&	0.512&	0.415&	0.643&	0.753&	0.534&	0.569&	0.512  \\ \cline{2-27} 
\multicolumn{1}{c|}{}                                    & \multicolumn{1}{c|}{}                                                                                   & Web-View                                                                            &0.356&	0.625&	0.751&	0.498&	0.538&	0.484&	0.341&	0.607&	0.797&	0.482&	0.564&	0.556&	0.439&	0.702&	0.806&	0.578&	0.612&	0.551&	0.423&	0.689&	0.797&	0.565&	0.600&	0.538  \\
\multicolumn{1}{c|}{}                                    & \multicolumn{1}{c|}{}                                                                                   & Call-Log                                                                            &0.427&	0.690&	0.798&	0.567&	0.602&	0.550&	0.413&	0.692&	0.750&	0.504&	0.595&	0.549&	0.414&	0.676&	0.786&	0.553&	0.588&	0.526&	0.385&	0.654&	0.769&	0.527&	0.564&	0.500  \\
\multicolumn{1}{c|}{}                                    & \multicolumn{1}{c|}{}                                                                                   & Benefits                                                                            &0.436&	0.702&	0.801&	0.558&	0.610&	0.560&	0.391&	0.653&	0.754&	0.504&	0.613&	0.543&	0.399&	0.662&	0.773&	0.538&	0.574&	0.512&	0.401&	0.659&	0.770&	0.538&	0.574&	0.512  \\
\multicolumn{1}{c|}{}                                    & \multicolumn{1}{c|}{\multirow{-4}{*}{\textbf{PoI}}}                                                     & Shopping                                                                            &0.429&	0.689&	0.785&	0.567&	0.598&	0.550&	0.418&	0.705&	0.755&	0.508&	0.593&	0.548&	0.380&	0.645&	0.761&	0.520&	0.557&	0.493&	0.386&	0.643&	0.756&	0.521&	0.558&	0.496  \\ \cline{2-27} 
\multicolumn{1}{c|}{}                                    & \multicolumn{1}{c|}{}                                                                                   & Web-View                                                                            &0.518&	0.768&	0.881&	0.644&	0.681&	0.626&	0.607&	0.764&	0.881&	0.632&	0.652&	0.604&	0.552&	0.753&	0.850&	0.654&	0.685&	0.635&	0.549&	0.758&	0.851&	0.654&	0.685&	0.633  \\
\multicolumn{1}{c|}{}                                    & \multicolumn{1}{c|}{}                                                                                   & Call-Log                                                                            &0.571&	0.813&	0.899&	0.694&	0.722&	0.672&	0.532&	0.824&	0.872&	0.708&	0.732&	0.687&	0.522&	0.756&	0.867&	0.640&	0.676&	0.617&	0.520&	0.764&	0.875&	0.643&	0.679&	0.618  \\
\multicolumn{1}{c|}{}                                    & \multicolumn{1}{c|}{}                                                                                   & PoI                                                                                 &0.591&	0.821&	0.903&	0.709&	0.735&	0.688&	0.587&	0.809&	0.907&	0.685&	0.741&	0.681&	0.533&	0.771&	0.875&	0.654&	0.688&	0.630&	0.544&	0.771&	0.877&	0.659&	0.694&	0.637  \\
\multicolumn{1}{c|}{}                                    & \multicolumn{1}{c|}{\multirow{-4}{*}{\textbf{Benefits}}}                                                & Shopping                                                                            & 0.570&	0.811&	0.898&	0.694&	0.722&	0.673&	0.550&	0.814&	0.906&	0.641&	0.696&	0.671&	0.471&	0.729&	0.847&	0.601&	0.640&	0.575&	0.469&	0.752&	0.876&	0.611&	0.652&	0.582 \\ \cline{2-27} 
\multicolumn{1}{c|}{}                                    & \multicolumn{1}{c|}{}                                                                                   & Web-View                                                                            & 0.162&	0.383&	0.538&	0.280&	0.330&	0.286&	0.139&	0.295&	0.465&	0.270&	0.302&	0.261&	0.206&	0.456&	0.597&	0.335&	0.380&	0.313&	0.237&	0.485&	0.628&	0.367&	0.414&	0.347 \\
\multicolumn{1}{c|}{}                                    & \multicolumn{1}{c|}{}                                                                                   & Call-Log                                                                            &   0.186&	0.425&	0.573&	0.308&	0.356&	0.307&	0.169&	0.399&	0.547&	0.266&	0.338&	0.291&	0.204&	0.429&	0.559&	0.320&	0.362&	0.301&	0.210&	0.447&	0.579&	0.332&	0.375&	0.312 \\
\multicolumn{1}{c|}{}                                    & \multicolumn{1}{c|}{}                                                                                   & PoI                                                                                 &      0.148&	0.390&	0.545&	0.272&	0.322&	0.272&	0.164&	0.307&	0.545&	0.271&	0.317&	0.291&	0.195&	0.426&	0.572&	0.314&	0.361&	0.296&	0.174&	0.317&	0.657&	0.247&	0.353&	0.266\\
\multicolumn{1}{c|}{\multirow{-20}{*}{\textbf{Telecom}}} & \multicolumn{1}{c|}{\multirow{-4}{*}{\textbf{Shopping}}}                                                & Benefits                                                                            &     0.228&	0.484&	0.630&	0.362&	0.409&	0.358&	0.226&	0.440&	0.556&	0.310&	0.391&	0.341&	0.189&	0.430&	0.573&	0.312&	0.358&	0.292&	0.186&	0.421&	0.569&	0.306&	0.353&	0.287   \\ 
\bottomrule
\end{tabular}
\end{table*}

\section{Details of Pairwise CDSR model} \label{section: pairwise}
Existing CDSR models predominantly focus on learning from only two domains at a time. 
In Table \ref{tab:result_1} and \ref{tab:result_2}, we evaluate the performance of these models on two-domain pairs and report their averaged performance across all pairs as the overall performance for each domain. 
Due to space limitations, the performance of individual domain pairs was not included in the main body of the paper (Tables \ref{tab:result_1} and \ref{tab:result_2}). 
Therefore, this section provides a detailed report of the performance for each domain pair, including the following CDR models: BiGCF~\cite{liu2020cross}, DTCDR~\cite{zhu2019dtcdr}, CMF~\cite{singh2008relational}, and DeepAPF~\cite{yan2019deepapf}, as well as the following CDSR models: C$^{2}$DSR~\cite{cao2022contrastive}, MAN~\cite{lin2023mixed}, $\pi$-Net~\cite{ma2019pi}, and MIFN~\cite{ma2022mixed}, which were not included in Tables \ref{tab:result_1} and \ref{tab:result_2}.
In the Amazon dataset used for CDR baselines, using the $Games$ domain as the source domain generally improved recommendation performance in the $Books$ domain when it was the target domain. A similar pattern was observed when the target domains were $Games$ and $Outdoors$, with $Toys$ and $Games$ serving as the respective source domains.
In the Telecom dataset, the target domain’s recommendation performance was maximized across most baseline models when the combinations of target and source domains were $Web$-$View$ and $Benefits$ or $Benefits$ and $Shopping$.
The results for CDSR baselines are as follows: In the Amazon dataset, the combinations of target and source domains—$Books$ and $Jewelry$, as well as $Outdoors$ and $Jewelry$—consistently yielded high recommendation performance across multiple baseline models. Similarly, in the Telecom dataset, the combinations of $Web$-$View$ and $Call$-$Log$, as well as $Benefits$ and $Web$-$View$, demonstrated strong performance for the target domain across various models.
This detailed analysis highlights the distinct relationships between the domains and provides a deeper insight into the cross-domain interactions.

\section{Extended Sensitivity Analysis Across All Domains} \label{section: extended_sensitivity}
\begin{figure*}[t]
    \centering
    \includegraphics[width=0.98 \linewidth]{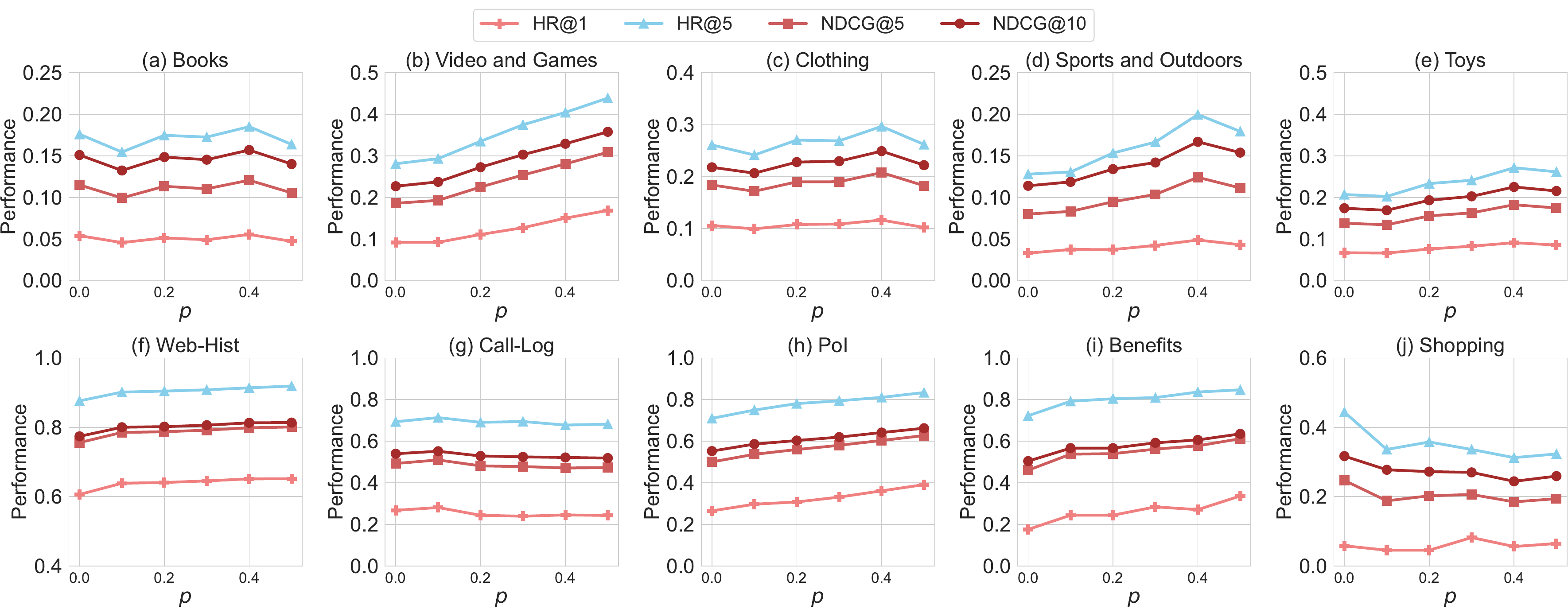}
    \caption{Domain-Wise Performance  Sensitivity analysis of $p$ for negative sampling
    }
    \label{fig:appendix_sensi}
    \end{figure*}

We conducted a sensitivity analysis on the hyperparameter $p$, which controls the proportion of cross-domain negative samples during contrastive learning.
For completeness, we report the results for all domains (Amazon), including those omitted in the main script, as shown in Fig.~\ref{fig:appendix_sensi}.
As discussed in the main script, we observe consistent performance improvements with increasing $p$ across most domains, although \textit{Call}-\textit{Log} and \textit{Shopping} domains did not exhibit a clear upward trend, whereas the others showed a robust increase pattern.

\section{Extended Analysis on Cold \& Warm Scenario Across All Domains} \label{section: extended_cold_and_warm}
\begin{figure*}[t]
    \centering
    \includegraphics[width=0.98 \linewidth]{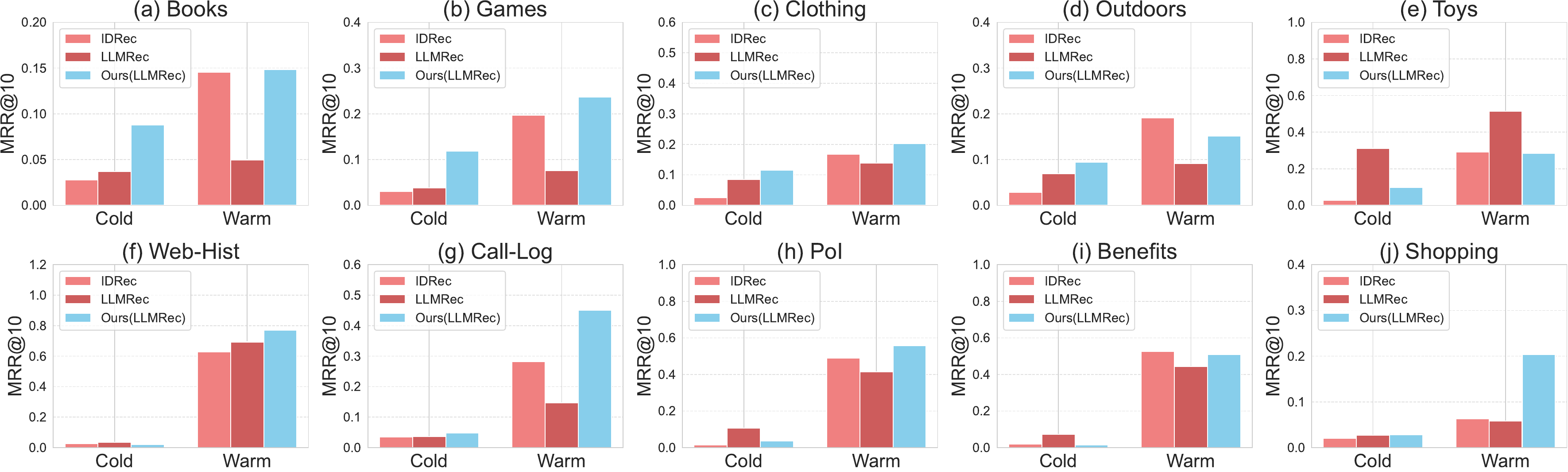}
    \caption{Domain-Wise Performance Comparison under Cold/Warm-Item Scenario
    }
    \label{fig:appendix_cold_warm_item}
    \end{figure*}

\noindent\textbf{Cold/Warm Item Scenario}
While Section~\ref{subsection:diss_cold_warm} presents the evaluation results on two representative domains due to space constraints, we report here the results across all domains under the cold- and warm-item scenarios. 
As shown in Fig.~\ref{fig:appendix_cold_warm_item}, the proposed model demonstrates its ability to capture item-level collaborative signals, achieving superior performance over IDRec models for both cold and warm items.

\begin{figure*}[t]
    \centering
    \includegraphics[width=0.98 \linewidth]{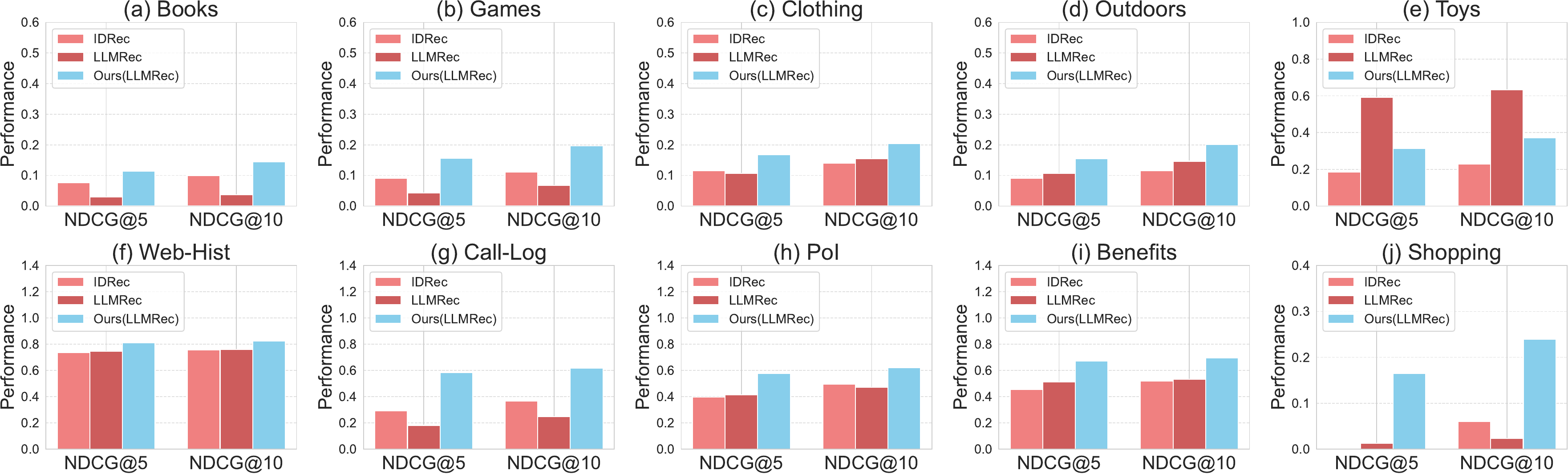}
    \caption{Domain-Wise Performance Comparison under Cold-User Scenario
    }
    \label{fig:appendix_cold_user}
    \end{figure*}

\noindent\textbf{Cold User Scenario}
To complement the findings in Section~\ref{subsection:diss_cold_warm}, which are limited to two domains due to space constraints, we present comprehensive results across all domains, as shown in Fig.~\ref{fig:appendix_cold_user}. 
DuELRec maintains consistent superiority over IDRec, with particularly substantial gains in domains where cold-user performance is hindered by limited collaborative signals.

\section{Extended Analysis on Instruction-Tuning Across All Domains} \label{section: extended_inst_tuning}
\begin{figure*}[t]
    \centering
    \includegraphics[width=0.98 \linewidth]{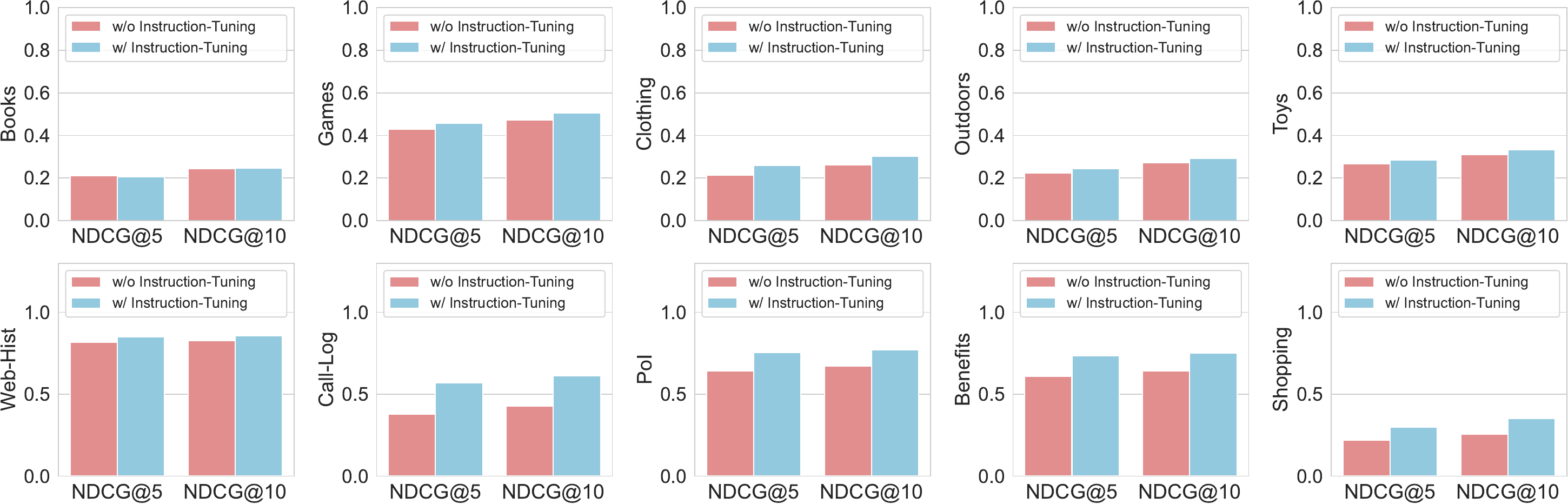}
    \caption{Domain-Wise Performance Comparison: Without vs. With Instruction-Tuning
    }
    \label{fig:instructino_tuning}
    \end{figure*}

In Section \ref{subsection:sft}, we proposed an auxiliary task that integrates collaborative signals with recommendation instructions to enable LLMRec to better adapt to textual user behavior sequences. 
This task involves fine-tuning the LLM using PEFT methods such as LoRA and prompt-tuning. 
We conducted experiments to compare performance with and without this auxiliary task.
As shown in Fig. \ref{fig:instructino_tuning}, the auxiliary task with PEFT significantly improved performance in all domains. 
This shows that allowing the LLM to learn from user behavior logs effectively improves recommendation performance.

\begin{table}[] \footnotesize 
\caption{Comparison of the HR@1 performance between retrieval-based (Ours) and generative model (A-LLMRec).}
\label{tab:generative_llm}
    \renewcommand{\arraystretch}{0.9}
\begin{adjustbox}{width=\columnwidth,center}
\begin{tabular}{cc|c|c|c}
\toprule
\multicolumn{2}{c|}{\textbf{Model}}                                        & \multirow{2}{*}{\textbf{\begin{tabular}[c]{@{}c@{}}Retrieval\\ -Based\end{tabular}}} & \multirow{2}{*}{\textbf{Generative}} & \multirow{2}{*}{\textbf{Gap(\%)}} \\ \cline{1-2}
\multicolumn{1}{c|}{\textbf{Dataset}}                  & \textbf{Domain}   &                                                                                      &                                      &                                   \\ \hline
\multicolumn{1}{c|}{\multirow{5}{*}{\textbf{Amazon}}}  & \textbf{Books}    &0.1096                                                                                      &  -                                    &  -                                 \\
\multicolumn{1}{c|}{}                                  & \textbf{Games}    &0.2941                                                                                      &    0.0730                                  &    -302.88                               \\
\multicolumn{1}{c|}{}                                  & \textbf{Jewelry}  &0.1487                                                                                      &0.0762                                      & -95.14                                  \\
\multicolumn{1}{c|}{}                                  & \textbf{Outdoors} &0.1223                                                                                      & 0.0639                                     &     -91.39                              \\
\multicolumn{1}{c|}{}                                  & \textbf{Toys}     &0.1624                                                                                      & 0.1138                                     &          -42.71                         \\ \hline
\multicolumn{1}{c|}{\multirow{5}{*}{\textbf{Telecom}}} & \textbf{Web-View} & 0.7219                                                                                     &           0.6840                           &         -5.54                          \\
\multicolumn{1}{c|}{}                                  & \textbf{Call-Log} &    0.3301                                                                                  &        0.1800                              &                -83.59                   \\
\multicolumn{1}{c|}{}                                  & \textbf{PoI}      &    0.5685                                                                                  &             0.2000                         &                   -184.82                \\
\multicolumn{1}{c|}{}                                  & \textbf{Benefits} &          0.4653                                                                            &  -                                    &     -                              \\
\multicolumn{1}{c|}{}                                  & \textbf{Shopping} &          0.1632                                                                            &                  0.0270                    &                            -504.44       \\ 
\bottomrule

\end{tabular}
\end{adjustbox}
  \begin{tablenotes}
    \item[*] * The - symbol indicates cases where performance evaluation was not performed due to our limited computing resources, or where performance was too low due to problems such as the hallucination problem in Generative LLMRec.
    \end{tablenotes}
\end{table}

\section{Future Work} \label{section: future_work}

In LLMRec, there are two primary approaches: (1) retrieval-based and (2) generative. 
The retrieval-based approach operates by recommending items strictly from a predefined set, ensuring that all recommendations are confined to this set. 
This makes it particularly advantageous when dealing with a large-scale item pool, as it eliminates the risk of hallucination. 
All the baselines introduced in Section \ref{section:experiments} (Experiments), including our proposed model, adopt this retrieval-based strategy.

On the other hand, the generative approach involves the LLM generating textual titles (or descriptions) of recommended items, typically selected from a relatively small candidate set (around 100 items). 
In this method, the candidate items are explicitly provided to the LLM as part of the prompt, along with instructions to identify the most suitable item from the given candidates. 
This approach holds significant promise for building personalized recommendation agents. However, it comes with a potential drawback: the risk of hallucination, where the model may generate recommendations for items that are not part of the predefined candidate set.

To evaluate the performance of generative LLMRec, we conducted experiments using a representative generative model, \textbf{A-LLMRec}~\cite{kim2024large}. 
A-LLMRec introduces a two-stage framework that aligns collaborative filtering models with LLMs, leveraging an alignment network to transfer collaborative knowledge into the token space of LLMs without requiring fine-tuning of either component. 
During the validation phase, this model generates the title of the \textit{top}-1 item among the 100 candidates included in the prompt.
Note that the candidate set in the prompt is composed of one positive sample and 99 negative samples, following the same experimental setting used during the validation phase in our study.

Table \ref{tab:generative_llm} presents a comparison of the HR@1 performance between A-LLMRec and our proposed model.
While its performance was lower compared to our retrieval-based model, it demonstrated reasonable effectiveness, even with the inherent risk of hallucination due to direct item generation.
These findings highlight the potential of generative LLMRec for future research. 
Our next goal is to develop generative methods capable of recommending items from extensive item pools without hallucination. 
We believe that such advancements in generative LLMRec could play a crucial role in creating highly personalized and reliable recommendation agents.

\end{document}